\documentclass[aps,twocolumn,floatfix]{revtex4}

\usepackage{amssymb}
\usepackage{amsmath}
 \usepackage{color}
\usepackage{psfrag}
\usepackage{ifpdf}
\ifpdf
\usepackage{epstopdf}   
\fi
\usepackage{mathtools}

\newcommand{\ket}[1]{\ensuremath{\vert#1\rangle}}
 \newcommand{\bra}[1]{\ensuremath{\langle #1\vert}}

\newtheorem{theorem}{Theorem}
\newtheorem{defin}{Definition}
\newtheorem{lem}{Lemma}
\newtheorem{corollary}{Corollary}

\newtheorem{claim}{Claim}

\newcommand{\im}{\mathrm{im}}

\newcommand{\wt}{\ensuremath{\mathrm{wt}}}

\newcommand{\colsupp}{\ensuremath{\mathrm{colsupp}}}
\newcommand{\rowsupp}{\ensuremath{\mathrm{rowsupp}}}

\def\id{{\mathchoice {\rm 1\mskip-4mu l} {\rm 1\mskip-4mu l} {\rm 1\mskip-4.5mu l} {\rm 1\mskip-5mu l}}}

\begin{document}

\title{A theory of single-shot error correction for adversarial noise}

\author{Earl T.\ Campbell}
\affiliation{Department of Physics \& Astronomy, University of Sheffield, Sheffield, S3 7RH, United Kingdom.}

\begin{abstract}
Single-shot error correction is a technique for correcting physical errors using only a single round of noisy check measurements, such that any residual noise affects a small number of qubits.   We propose a general theory of single-shot error correction and establish a sufficient condition called good soundness of the code's measurement checks.  Good code soundness in topological (or LDPC) codes is shown to entail a macroscopic energy barrier for the associated Hamiltonian.  Consequently,  2D topological codes with local checks can not have good soundness.  In tension with this, we also show that for any code a specific choice of measurement checks does exist that provides good soundness.  In other words, every code can perform single-shot error correction but the required checks may be nonlocal and act on many qubits.   If we desire codes with both good soundness and simple measurement checks (the LDPC property) then careful constructions are needed.  Finally, we use a double application of the homological product to construct quantum LDPC codes with single-shot error correcting capabilities.  Our double homological product codes exploit redundancy in measurements checks through a process we call metachecking.
\end{abstract}

\maketitle

In the simplest model of quantum error correction,  noise affecting qubits is corrected under the assumption that measurements  are performed perfectly.   In reality, measurement results will be unreliable.  The standard tactic for combating measurement noise is to repeat the measurements and build a timeline of measurement data.  Error correction software can then attempt to infer the most likely explanation of the observed measurement results.   The number of measurement rounds required will typically grow with the code size.  Recently, single-shot error correction was proposed by Bombin as a radically different solution to measurement noise~\cite{bombin2015single}.   In single-shot error correction, no repeated measurements are needed.  Benefits include faster error correction and an inherent resilience to temporally correlated noise~\cite{bombin2016resilience}.  However, few codes are known to support single-shot error correction.  This idea was proposed in the setting of topological codes in three or four spatial dimensions, such as the three dimensional gauge colour code~\cite{bombin15} and four dimensional toric code~\cite{dennis02}.  Very recently, it has been reported that quantum expander codes also allow for single-shot error correction~\cite{leverrier18}.  Quantum data-syndrome codes are also closely related to single-shot codes~\cite{fujiwara14,ashikhmin16}. The development and implementation of decoding algorithms for single-shot error correction is also limited with only a few examples~\cite{brown15,breuckmann2016local}. So far progress has been focused on specific examples and one of our goals here is to lay down a common framework within which single-shot error correction can be understood and analysed. 

In the idealised setting of perfect measurements, error correction will return the system back into the code-space, either with or without a logical error. A quantum code is parametrised by its distance $d$ where perfect measurements can always detect noise on fewer than $d$ qubits.  Consequently, noise on any $(d-1)/2$ qubits can be successfully corrected, even if the damaged qubits are chosen by an adversary who is attempting to corrupt the quantum information. Here we consider adversarial noise in the single-shot setting.  We allow for physical qubit errors and measurement errors to appear in any pattern but affecting a limited number of qubits and measurements.  Given corrupt measurement data, error correction may not even return the system to the code-space, but will leave some residual qubit error.  Single-shot error correction aims to control the size of this residual error.   Central to achieving this is the notion of soundness~\cite{aharonov2015quantum,hastings2016quantum}. Loosely, a code has good soundness if small measurement syndromes can be produced by small qubit errors.  Good soundness is closely related to local testability of codes~\cite{aharonov2015quantum,hastings2016quantum} and energy barriers in self-correcting quantum memories~\cite{alicki2010thermal,terhal2015review,brown2016review}.  It is clear that the 4D toric codes have good soundness properties.  We shall also show that good soundness entails the existence of a macroscopic energy barrier~\cite{Bacon06,alicki2010thermal} and consequently 2D topological codes cannot possess good soundness properties.  However, we also show that given any quantum code we can adapt the check measurements to ensure good soundness, though in the process any topological and or low-density parity check (LDPC) properties will be lost.  This leads to the surprising insight that any quantum error correction code can perform single-shot error correction, provided we are content with error correction measurements involving a large number of qubits.  The interesting challenge is then to find codes that combine good soundness with LDPC properties. 

The second part of this work provides techniques for constructing quantum codes with good single-shot correcting capabilities.  Our approach is to use a double application of the homological, or hypergraph, product.  The hypergraph product was first used by Tillich and Z\'{e}mor~\cite{tillich2014quantum} to show that any two classical codes can be combined to make a new quantum code. Unlike the standard CSS construction, no special relationship between the two codes is required by the hypergraph product.   If the original classical codes are good LDPC codes (constant rate and linear distance) then the hypergraph product produces a quantum LDPC code with a good rate  (the number of logical qubits $k$ scaling as a constant fraction of the number of physical qubits $n$) and distance scaling as $\Theta(\sqrt{n})$;  becoming the first quantum LDPC code to achieve such parameters.  Subsequently, Leverrier, Tillich and Z\'{e}mor proposed quantum expander codes that result from the hypergraph product of two expander graph codes, which are a specific family of good LDPC codes.  The expansion properties of these codes enabled them to devise an efficient decoder correcting adversarial errors affecting upto $O(\sqrt{n})$ qubits.  Later it was shown that the decoder could correct $O(n)$ random errors with high probability~\cite{fawzi2018} and support single-shot error correction~\cite{leverrier18}.   Furthermore, maximum likelihood decoding has been investigated for quantum expander codes providing both analytical lower bounds~\cite{dumer2015thresholds} and numerical estimates~\cite{kovalev2018numerical}.

Our approach here has overlap in the formal techniques but is more widely applicable since it does not depend on the strong assumption that the initial classical code is an expander graph code.  The hypergraph product is closely related to the homological product used in the study of algebraic topology.  Bravyi and Hastings~\cite{bravyi2014homological} used the homological product to construct codes with a linear distance and good rate;  though they were not strictly low-density parity check codes.  Audoux and Couvreur studied repeated application of the homological product~\cite{audoux2015tensor}.  

We will use two applications of the homological product to design single-shot codes from any classical code.   Two applications of the homological product generates a structure that in homology theory would be described as length 4.  Sometimes this length-4 algebraic structure can be embedded within a geometrically local 4-dimensional manifold and the resulting quantum code would be a 4-dimensional topological code.  Given a family of LDPC classical codes, our construction gives a family of LDPC quantum codes with good soundness, successfully combining these two desirable properties. However, our approach is inherently algebraic, providing many codes with no natural spatial topology, unless the original classical codes are topological.  From the perspective of practical implementations, a topological code of modest dimension may seem preferable.  However, topological codes are constrained by trade-off bounds on the achievable code parameters~\cite{bravyi2010tradeoffs,delfosse2013tradeoffs} and so non-topological codes can be much more efficient.  

We begin by reviewing the key concepts (Sec.~\ref{Sec_Intro}) before giving a more technical statement of the main results (Sec.~\ref{Sec_summary}).  We prove sufficient conditions for single-shot error correction in Sec.~\ref{Sec_Conditions}.  We discuss the relationship between soundness and energy barriers in Sec.~\ref{Sec_Energy}.  We show how measurement checks can be redefined for any code to provide good soundness in Sec.~\ref{Sec_Simple}. We give a general overview of how homology theory can be used to describe quantum codes in Sec.~\ref{Sec_Homology_theory}.  This establishes the technical groundwork for Sec.~\ref{Sec_Constructions} where we give code constructions that meet our criteria using a double application of the homological product.  We conclude with a discussion of the remaining open problems and the limitations of considering adversarial noise rather than stochastic noise.

\section{Key concepts}
\label{Sec_Intro}

The preliminary material covered in this section draws from the work of Bombin~\cite{bombin2015single,bombin2016resilience} and was influenced by Brueckmann's thesis~\cite{breuckmann2018phd}, though our presentation is less topological and has some new ideas. 

\subsection{Stabiliser codes}

An $n$ qubit error correcting code storing $k$ logical qubits can be represented by a projector $\Pi$ onto the codespace.  Stabiliser codes are an important class where $\Pi$ can be described in terms of the code stabiliser $\mathcal{S}$.  That is, $\mathcal{S}$ is an abelian subgroup of the Pauli group such that for all $S \in \mathcal{S}$ we have $S \Pi = \Pi  S = \Pi$.  To perform error correction we measure some set of checks $\mathcal{M} \subset \mathcal{S}$ that generate $\mathcal{S}$ under multiplication. We require that $\mathcal{M}$ suffices to the generate the whole stabiliser of the codespace but we allow for the possibly of $\mathcal{M}$ being overcomplete.  We define the weight $\mathrm{wt}(\cdot)$ of a Pauli operator $P$ as the number of qubits on which $P$ acts nontrivially (the identity is the only trivial Pauli).  Given a family of check sets $\mathcal{M}_n$ with index $n$, which we will call a \textit{check family}, we find there is a corresponding code family $\Pi_n$.  For a given code family, there may be many different choices of check family, so many statements are more precisely defined with respect to check families.  For instance, we have a notion of low-density party check (LDPC) and we say a check family is LDPC if there exists a constant $C$ such that for every $n$ 
\begin{enumerate}
	\item For all $S \in \mathcal{M}_n$ we have $\mathrm{wt}(S)\leq C$;
	\item For every physical qubit in the code, there are no more than $C$ checks in $\mathcal{M}_n$ that act non-trivially on that qubit.
\end{enumerate}
It is crucial that the constant $C$ is the same for every member of the family.  One practical consequence is that for codes with an LDPC check family, the complexity of measuring checks does not increase with the code size. Crudely, one can say a code family is LDPC if there exists at least one corresponding LDPC check family. Note that topological code families are always LDPC.

Also important is the code distance $d_Q$. We use the subscript $Q$ to distinguish this from the single-shot distance (denoted $d_{ss}$) that we define later. The distance $d_Q$ is simply the minimum $\mathrm{wt}(P)$ over all $P$ such that $P \Pi = \Pi  P$ but $P \notin \mathcal{S}$.  It is useful to also define the min-weight $\wt_{\mathrm{min}}$ of a Pauli operator, which is 
\begin{equation}
	 \wt_{\mathrm{min}} (P) := \{  \mathrm{wt} (PS) : S \in \mathcal{S}  \}.
\end{equation}
To summarise,  an $[[n,k,d_Q ]]$ code has parameters $n$ (number of physical qubits), $k$ (number of logical qubits) and $d_Q$ (qubit code distance).  

The measurement syndrome is the result of measuring $\mathcal{M}=(M_1, M_2, \ldots , M_m)$.  Given a physical Pauli error $E$ we can denote $\sigma(E)$ as the syndrome due to $E$ assuming perfect measurements.  We use the convention that $\sigma(E)$ is a binary column vector with elements
\begin{equation}
	\label{Eq_syndrome_def}
	[\sigma(E)]_i = \begin{cases}
	1 & \mbox{ if } E M_i = - M_i  E \\
	0 & \mbox{ if } E M_i =  M_i  E \\
	\end{cases}
\end{equation}
We will be interested in the weight of the syndrome and always use $| \ldots |$ to denote the Hamming weight of binary vectors.  The Hamming weight is the number of nonzero elements.  

\subsection{Single-shot error correction}
\label{Intro_soundness}

A decoder is an algorithm that takes a measurement syndrome $s \in \mathbb{Z}_2^m$ and outputs a recovery Pauli operator $E_{\mathrm{rec}}$.   We model measurement errors as introducing an additional syndrome vector $u$ so that we physically observe syndrome $s=\sigma(E)+u$ where $E$ is the physical error.  Good decoder design would ensure that given $s$ the recovery is such that residual error $E_{\mathrm{rec}}  E$ has low min-weight.   We propose the following definition
\begin{defin}[Single-shot error correction]
	\label{single_shot_def}
	Let $p$ and $q$ be integers and $f : \mathbb{Z} \rightarrow \mathbb{R}$ be some function with $f(0)=0$.  We say a check set is $(p, q, f)$ single-shot if there exists a decoder such that for all $u$ and $E$ such that
	\begin{enumerate}
		\item  $|u| < p$ ; and
		\item  $f(2|u|) + \mathrm{wt}(E) < q $ 
	\end{enumerate}
 the decoder takes syndrome $s=\sigma(E)+u$ and outputs recovery operation  $E_{\mathrm{rec}}$ such that  $\wt_{\mathrm{min}}(E_{\mathrm{rec}} \cdot E) \leq f(2 |u|)$.
\end{defin}
This captures all instances of single-shot error correction known to the author.  We are interested in good cases where $p$ and $q$ are large and $f$ is in some sense small.  A very bad case is when $p=1$ so that no measurement errors ($|u|<1$) can be tolerated. A more rigorous notion of good single-shot properties requires us to consider not just a single instance but an infinite check-family.
\begin{defin}[Good single-shot families] \label{GoodssDEF}
	Consider an infinite check family $\mathcal{M}_n$ of $n$-qubit codes.  We say the family is a good single-shot family if each $\mathcal{M}_n$ is $(p ,q , f)$ single-shot where
	\begin{enumerate}
		\item $p$ and $q$ grow with $n$ such that $p,q \geq a n^b$ for some positive constants $a,b$.  That is, $p,q \in \Omega( n^b )$ with $b>0$;
		\item  and $f(x)$ is some polynomial that is monotonically increasing with $x$ and independent of $n$.
	\end{enumerate}  
\end{defin}
We need $p$ and $q$ to grow so that we can tolerate more errors as the code size grows.  We want $f$ to be independent of $n$ so that the residual errors remain contained.

Single-shot error correction is defined for a single round but it is informative to see what the consequences are for $N$ rounds of error correction.  We use a label $\tau \in \{ 1, \ldots , N \}$ for the round number.  On round $\tau$, we denote $u_{\tau}$ for the measurement errors and $E_{\tau}$ for the new physical errors.  We must combine $E_{\tau}$ with the residual error from the previous round $R_{\tau-1}$ to obtain the total error $E_{\tau}R_{\tau-1}$.  For the $\tau^{\mathrm{th}}$ round to satisfy the conditions in Def.~\ref{single_shot_def} we need that $|u_{\tau}| < p$  and
\begin{equation}
			f(2|u_{\tau}|) + \mathrm{wt}(E_{\tau} R_{\tau-1} ) < q .
\end{equation}	
Assuming similar conditions were satisfied on the previous round, we may upper bound $\mathrm{wt}( R_{\tau-1} )$ using Def.~\ref{single_shot_def} and have
\begin{equation}
	f(2|u_{\tau}|) + f(2|u_{\tau-1}|) + \mathrm{wt}(E_{\tau})  < q .
\end{equation}	
Therefore, provided the measurement errors and new physical errors are small for every round, the residual error will be kept under control over many rounds and not grow in size.

The above definition of single-shot error correction is difficult to analyse since it contains the clause ``if there exists a decoder" and there are many possible decoders.  Therefore, we also consider a complementary concept called soundness which will be shown to entail single-shot error correction.  Roughly, this extra property is that for low weight syndromes there exists a low weight physical error producing the syndrome.  More formally,
\begin{defin}[Soundness] \label{RoughssDEF}
	Let $t$ be an integer and $f : \mathbb{Z} \rightarrow \mathbb{R}$ be some function called the soundness function with $f(0)=0$.  Given some set of Pauli checks $\mathcal{M}$, we say it is $( t, f)$-sound if for all Pauli errors $E$ with $| \sigma(E) | = x < t$,  it follows that there exists an $E^\star$ with $\sigma(E^\star) = \sigma(E)$ such that  $\wt(E^\star) \leq f(x)$.
\end{defin}
The phrase soundness comes from the literature on locally testable codes~\cite{aharonov2015quantum,hastings2016quantum}.  In particular, the above definition is similar to Def 14 of Ref.~\cite{aharonov2015quantum} though this earlier work did not allow for the $| \sigma(E) |  < t$ clause.  

Again, good soundness would mean ``small" $f$.   More rigorously, we define the following notion of goodness
\begin{defin}[Good soundness] \label{GoodssDEF}
  Consider an infinite check family $\mathcal{M}_n$.  We say the family has good soundness if each $\mathcal{M}_n$ is $(t, f)$-sound where:
  \begin{enumerate}
  	  \item $t$ grows with $n$ such that $t \geq a n^b$ for some positive constants $a,b$.  That is, $t \in \Omega( n^b )$ with $b>0$;
  	  \item  and $f(x)$ is some polynomial that is monotonically increasing with $x$ and independent of $n$.
  	\end{enumerate}  
\end{defin}
The intuition behind $f$ being a polynomial is that we are formalising an algebraic version of an area or volume law that is encountered in topological codes.  For instance, in the classical 2D Ising model we know that the area within a boundary follows a quadratic scaling (you may wish to look ahead to Fig.~\ref{Fig_ToricIsing}b3).  Ultimately, $f$ will govern the size of residual errors after performing single-shot error correction, so we do not want it to grow with the number of qubits.  In contrast, $t$ captures the scale at which this boundary law breaks down and so it must grow with the code size to enable single-shot error correction of larger errors as the code grows.

 It is clear that not all check families have good soundness.  For 2D toric codes with the standard choice of checks, an error violating only 2 checks can be of arbitrarily large size.

\subsection{Energy barriers}
\label{Intro_energy}

Energy barriers play an important role in the design of passive quantum memories~\cite{brown2016review,terhal2015review}.  While passive quantum memories are a distinct topic from active single-shot error correction, the two topics are intertwined.  Earlier work~\cite{aharonov2015quantum} has commented on the relationship between soundness and energy barriers, though they used a more restrictive notion of soundness.   For a stabiliser code with checks $\mathcal{M}$ we define a Hamiltonian 
\begin{equation}
H =- \sum_{S \in \mathcal{M}} S .
\end{equation}	
We are interested in walks of quantum states $W=\{  \psi_0 ,  \psi_1 , \psi_2, \ldots , \psi_L  \}$ that fulfil 
\begin{enumerate}
	\item groundstates: $\psi_0 $ and $\psi_L$ are groundstates of $H$;
	\item orthogonality: $\psi_0 $ and $\psi_L$ are orthogonal;
	\item  local errors: for every $j \in [1,L]$ there exists a single-qubit Pauli $P_j$ such that $\ket{\psi_j}=P_j \ket{\psi_{j-1}}$.
\end{enumerate}	
For every such walk we associate an energy penalty
\begin{equation}
ep(W) = \mathrm{max}_{\psi_j \in W} \bra{\psi_j} H \ket{\psi_j}  - E_{gs} ,
\end{equation}	
where $E_{gs}$ is the ground state energy. The energy barrier of check set $\mathcal{M}$ and associated Hamiltonian is then the minimum $ep(W)$ over all walks $W$ satisfying the above conditions.  Less formally, the energy barrier is the minimum energy required to go from one ground state to another. 

Every quantum code will have some size energy barrier.  We are really interested in the scaling with code size.  Given an infinite check family $\mathcal{M}_n$ of $n$-qubit codes, if the energy barrier scales as $\Omega(n^c)$ for some positive constant $c$, then we say the family has a \textit{macroscopic} energy barrier.

\subsection{Measurement redundancy and single-shot distance}

We have allowed for some redundancy so that checks $\mathcal{M}$ may be overcomplete. This is pivotal for us to capture the single-shot properties of the 4D toric codes since they are only known to exhibit good soundness when an overcomplete set of checks are used.   We quantify the amount of redundancy in a measurement scheme as the ratio between the number of measurements performed and the minimum number required to generate the stabiliser of the code and use $\upsilon$ to denote this ratio.  Good soundness can always be achieved by allowing $\upsilon$ to grow with $n$ by simply repeating the same measurements. Rather, the most interesting cases are check families where $\upsilon$ is no more than a small constant factor.  There may also be interesting intermediate cases where $\upsilon$ grows but slowly (e.g. sublinearly), though a constant factor is more desirable and is what we prove later in our constructions.  Since topological codes can use redundancy to achieve good soundness, it is reasonable to ask whether redundancy is necessary for good soundness?   We will see later that redundancy is not always essential for good soundness (see Thm.~\ref{THM_soundness_simple} and Sec.~\ref{Sec_Simple}).  However, it seems that redundancy does play an important role when one attempts to marry good soundness with LDPC properties.

Check redundancy provides consistency conditions that one can inspect for evidence of measurement errors.   These are checks on checks and we call them metachecks.  They do not represent a physical measurement but classical postprocessing on the measurement outcomes.  It is essentially a classical error correcting code that can be represented by a parity check matrix $H$.  Given a binary string $s$ representing the outcome of syndrome measurements, we say $Hs$ is the metacheck syndrome, where $Hs$ is evaluated modulo 2.  If there are no measurement errors then $s=\sigma(E)$ where $E$ is the physical error.  Recall that we model measurement errors as introducing an additional error $u$ so that $s=\sigma(E)+u$.  Since the metachecks are intended to look for measurement errors, we require that $H \sigma (E) =0$ for all $E$.   It follows that the metasyndrome $H s = H(\sigma(E)+u)=Hu$ depends only on the measurement error $u$.   There will always exist a maximal set of metachecks $H_{\mathrm{max}}$ such that $H_{\mathrm{max}}s = 0$ if and only if there exists an error $E$ such that $s=\sigma(E)$.  However, we are flexible and allow for $H$ to contain fewer checks than $H_{\mathrm{max}}$, so that not all check redundancies are metachecked.  While it might seem odd to not use the maximum information present, this occurs naturally in some local decoders for topological codes where local metachecks are used but non-local metachecks are ignored by the decoder (see for instance the discussion on ``energy-barrier limited decoding" in Ref.~\cite{breuckmann2016local}). Given a non-maximal set of meta-checks, there are syndromes $s$  that pass all metachecks ($Hs=0$) and yet there is no error $E$ satisfying $ s=\sigma(E)$.   This motivates the following definition.
\begin{defin}[Single-shot distance]
	For a code with checks $\mathcal{M}$ and metacheck matrix $H$ we define the single-shot distance as
	\begin{equation}
	 d_{ss} = \mathrm{min} \{ | s | :  H s =0 ,  s \notin \mathrm{im}(\sigma)   \}.
	\end{equation}
	We use the convention that 	$d_{ss} = \infty$ if for all $s$ there exists some $E$ such that $s = \sigma(E)$.
\end{defin}
Here, $\mathrm{im}(\sigma)$ is the image of map $\sigma$, which is the set of $s$ such that $s=\sigma(E)$ for some $E$.  A equivalent definition is that $d_{ss}$ is the minimal weight $s$ such that $H s=0$ but $H_{\mathrm{max}}s \neq 0$.
The single-shot distance relates to how many measurement errors can be tolerated before a failure occurs that we call a metacheck failure.  In a metacheck failure, the syndrome has no explanation in terms of qubit errors.  

We remark that for any $\mathcal{M}$ we can always choose $H=H_{\mathrm{max}}$ and then $d_{ss}$ is infinite.  However, sometimes a finite single-shot distance may be preferred to ensure that the metacheck decoding process can be implemented using a local decoder~\cite{breuckmann2016local}.   For a code with metachecks we extend the notation $[[n,k,d_Q]]$ to $[[n,k,d_Q, d_{ss}]]$.

\section{Summary of results}
\label{Sec_summary}

Here we prove the following:
\begin{theorem}[Single-shot success]
	\label{THM_main}
	Consider a quantum error correcting code  with parameters $[[n,k,d_Q, d_{ss}]]$ and check set that is $( t, f)$-sound.   It is also $(p,q,f)$ single-shot where 
	\begin{align}
		p & = \frac{1}{2} \mathrm{min}[d_{ss}, t] \\
		q & = d_{Q}/2.
	\end{align}	
\end{theorem}
For the above bounds to be useful, the code must have a soundness function $f$ that is fairly gentle (e.g. some polynomial).  The proof is mostly linear algebra and is given in Sec.~\ref{Sec_Conditions}.

Our second result is an observation on the connection between soundness and energy barriers. 
\begin{theorem}
	\label{THM_soundness_implies}
Any LDPC check family with good soundness and code distance $d_Q$ growing as $\Omega(n^c)$ for some constant $0 < c$ will also have a macroscopic energy barrier.
\end{theorem} 
This is proved in Sec.~\ref{Sec_Energy}. We remark that Aharonov and Eldar made a similar observation~\cite{aharonov2015quantum} though using a much stronger notion of soundness.  Since Bravyi and Terhal proved that no 2D topological code can have a macroscopic energy barrier~\cite{bravyi2009no}, it follows immediately that
\begin{corollary}
\label{No_go_corollary}	Any 2D topological check family with code distance $d_Q$ growing as $\Omega(n^c)$ for some constant $0 < c$  will not have good soundness.
\end{corollary} 
We thank Michael Beverland for pointing out that this corollary follows directly from Thm.~\ref{THM_soundness_implies} and the Bravyi and Terhal result.  

Next, we show that
\begin{theorem}
	\label{THM_soundness_simple}
	For any $n$-qubit quantum error correcting code we can find a set of checks generating the code stabiliser (without any redundancy) such that these checks are $( \infty, f(x)=x)$-sound.
\end{theorem} 
The proof is elementary and given in Sec.~\ref{Sec_Simple}.  While this is a simple result, it carries important implications for our understanding of soundness.  It shows that any code family can be bestowed with good soundness by appropriate choice of checks, but in the process the LDPC property may be lost.  Therefore, the interesting question is for which code families we can find checks that are simultaneously LDPC and of good soundness. 

Our last main result is a recipe for quantum codes with the required properties.  We show that
\begin{theorem}[Construction of single-shot codes]
	\label{THM_construct}
	Given a classical error correcting code with parameters $[n,k,d]$ we can construct a quantum error correcting code with parameters $[[n_Q, k^4, d_Q \geq d, d_{ss}=\infty ]]$ where
	\begin{align}
		n_Q & = n^4 + 4 n^2 (n-k)^2 + (n-k)^4  .
	\end{align}
	 Furthermore, the resulting checks are $( d , f)$-sound and also $(\frac{d}{2} , \frac{d}{2} ,f)$ single-shot, with $f(x)=x^3 / 4$ or better. The check redundancy is bounded $\upsilon < 2$.  Given a classical LDPC check family, this construction gives a quantum LDPC check family.  Given a classical check family where $d \in \Omega(n^a)$ we have a good single shot family.
\end{theorem}
We remark that the distance bound $d_Q \geq d$ and soundness properties are loosely bounded.  Indeed, very recently Zeng and Pryadko~\cite{zeng2018higher} considered the same code family and showed that $d  = d^2 $.  

Before giving the proof of Thm.~\ref{THM_construct}, we establish how the mathematics of homology theory and chain complexes can be used to define quantum codes with metachecks.  As such, we provide a pedagogical interlude in Sec.~\ref{Sec_Homology_theory} that introduces this correspondence. The proof is then given in Sec.~\ref{Sec_Constructions} and uses the homological product on chain complexes.  Where possible we have converted abstract homological proofs into linear algebra.  The constructions of Thm.~\ref{THM_construct} will emerge as a simple, special case of the techniques explored  in Sec.~\ref{Sec_Constructions}, and we will see that codes with finite single-shot distance are also possible.  An important metric is the encoding rate, the number of logical qubits per physical qubit $k_Q/n_Q$.  The expressions for the inverse rate are neater to write
\begin{align}
		\frac{n_Q}{k_Q} & =   \frac{n^4 + 4 n^2 (n-k)^2 + (n-k)^4 }{k^4} \\ \nonumber
		& =  6 \left( \frac{n}{k} \right)^4 - 12 \left( \frac{n}{k} \right)^3   + 10 \left( \frac{n}{k} \right)^2 - 4 \left( \frac{n}{k} \right) + 1   .
\end{align}
From this, it is clear that for any family of classical codes with constant rate $n/k \leq A$, will yield a family of quantum codes with constant rate $n_Q/k_Q \leq A_Q \sim O( A^4 )$. 

\section{Conditions for successful  single-shot error correction}
\label{Sec_Conditions}

This section proves that soundness leads to single shot error correction as stated in Thm.~\ref{THM_main}.  Our analysis will use a minimum weight decoder defined as follows:
\begin{defin}[MW single-shot error decoding] \label{Def_SS_decode}
Given measurement outcomes $s = \sigma(E) +u$, a minimum weight decoder performs the following 2 steps
\begin{enumerate}
		\item Syndrome decode: find $s_{rec}$ with minimal $|s_{rec}|$ such that $s+s_{rec}$ passes all metachecks (so $H(s+s_{rec})=0$); 
		\item Qubit decode: find $E_{rec}$ with minimal  $\wt( E_{rec})$ such that $\sigma( E_{rec} )=s + s_{rec}$;
\end{enumerate}
We call $R= E \cdot E_{rec}$ the residual error.
\end{defin}
This is the most common notion of weight minimisation and for instance was suggested by Bombin~\cite{bombin2015single}.  Other decoders may correct more errors or may be more efficient to  implement.  However, the minimum weight decoder is especially useful in the following analysis.

Note that it is not possible to always find solutions to the above problem.  For instance, one may find a minimising $s_{rec}$ but then there is no $E_{rec}$ satisfying the second condition.  We call such an event a metacheck failure, but we do have the following guarantee 
\begin{lem}[Meta-check success]
We can find a solution to MW single-shot decoding provided that $|u| < d_{ss}/2$.  
\end{lem}
The proof is essentially the same as standard proofs for correcting adversarial qubit errors.  Metacheck failures correspond to cases where there exists a minimal weight $s_{rec}$ where $H(s+s_{rec})=0$ but there is no physical Pauli error $E$ such that $\sigma(E)=s+s_{rec}$.  Note that whenever we use ``$+$" between two binary vectors it should be read as addition modulo 2. First, we note that $H(s+s_{rec})=H(\sigma(E)+u+s_{rec})$ and using $H\sigma(E)=0$ we get that $s_{rec}$ must satisfy $H(u+s_{rec})=0$.  Since, $s_{rec}=u$ would satisfy this requirement and $s_{rec}$ is minimum weight, we infer that $|s_{rec}|\leq |u|$. Using the triangle inequality we get $|s_{rec}+u| \leq 2 |u| < d_{ss} $.  By the definition of single-shot distance, it follows that there exists a physical error $E'$ such that $\sigma(E')=s_{rec}+u$.  Using the syndrome relation $\sigma(E \cdot E') = \sigma (E)+ \sigma(E')$ we obtain
\begin{equation}
	\sigma(E \cdot E') = s +u + s_{rec}+u = s + s_{rec}.
\end{equation}
Therefore, there is always a physical error (e.g. $E_{rec}=E \cdot E'$) consistent with the repaired syndrome $s + s_{rec}$ and the lemma is proved.

The above proof shows that the code can tolerate up to $d_{ss}/2-1$  adversarial measurement errors and provide a solution to single-shot decoding.   However, the story is not finished since even if a metacheck failure does not occur, a conventional logical failure might yet occur.   Therefore, next we address the question of how we can ensure the residual error $R=E_{rec} \cdot E$ has bounded size.  From  $\sigma( E_{rec} )= s + s_{rec}$ we deduce $\sigma(R)=u+s_{rec}$ and so
\begin{equation}
\label{Eq_small_residual_synd}
   |\sigma(R)| \leq 2 |u| < d_{ss}
\end{equation}
This prompts the question, given a small syndrome (consistent with metachecks) does there even exists a small weight physical error generating this syndrome!  Indeed, this is not always the case; unless the code has nice soundness properties.  Using our notion of soundness we can prove the following
\begin{lem}[An upper bound on residual error]
	\label{ResErrLem}
	Consider a quantum error correcting code with parameters $[[n,k,d_Q, d_{ss}]]$ that is $( t, f)$-sound.  Given measurement error $u$ and physical error $E$.  If
	\begin{enumerate}
		\item  	$|u| < d_{ss}/2$ : the measurement error is small enough to ensure no metacheck failures;
		\item    $|u| < t/2$  :  the measurement error is small enough to use soundness properties; 
		\item    $f(2 |u|) + \wt(E)< d_Q/2$ : the combined errors are sufficiently small;
	\end{enumerate}	
It follows that a solution to MW single-shot decoding will yield a residual error $R=E \cdot E_{rec}$ with $\wt_{\mathrm{min}}( R ) \leq f( 2|u|  )$.
\end{lem}
We know from above (Eq.~\ref{Eq_small_residual_synd}) that the residual error $R$  satisfies $|\sigma(R)|\leq 2 |u| < d_{ss}$.  By using the definition of $(t,f)$-soundness, we know that provided $2 |u| < t$ there exists an $R^\star$ such that $\sigma (R)=\sigma (R^\star)$ and $\wt( R^{\star}) \leq f( 2 |u|  )$.	 It remains to show that $S=R  R^\star$ is a stabiliser of the code as this would entail that $\wt_{\mathrm{min}}( R ) \leq \wt ( R^\star ) \leq f( 2 |u|  )$.  Clearly, $\sigma(R R^\star)=\sigma(S)=0$ so $S$ is either a stabiliser or a nontrivial logical operator.  It can only be a nontrivial logical operator if  $ d_Q \leq \wt(R R^{\star}) $.  The rest of the proof shows that we instead have $\wt(R R^{\star}) < d_Q$ and so $S$ is a stabiliser.  We start with
\begin{align}
	R \cdot R^\star = E \cdot E_{rec} \cdot R^\star , 
\end{align}
and 
\begin{align}
  \wt( R \cdot R^\star ) = \wt( E \cdot E_{rec} \cdot R^\star ). 
\end{align}
Using the triangle inequality
\begin{align}
  \wt( R \cdot R^\star ) \leq  \wt( E_{rec} ) +\wt(   E \cdot  R^\star ) .
\end{align}
Since, $E_{rec}$ is a minimum weight solution, we can assume that $\wt( E_{rec} ) \leq \wt(   E \cdot  R^\star )$, and hence
\begin{align}
  \wt( R \cdot R^\star ) \leq 2 \wt( E \cdot  R^\star   ) \leq 2 \wt( E ) + 2 \wt(  R^\star   ).
\end{align}
Using again that $\wt( R^{\star}) \leq f( 2 |u|  )$ we obtain
\begin{align}
  \wt( R \cdot R^\star )\leq 2 ( f(2 |u|) + \wt( E ) ) .
\end{align}
We are interested in when the LHS is upper bounded by $d_Q$, which follows from the RHS being upper bounded by $d_Q$, which is precisely the third condition of the lemma. Therefore,  $\wt( R \cdot R^\star ) < d_Q$ and consequently $R=S \cdot R^\star$.  This proves the lemma, and Thm.~\ref{THM_main} follows by simply rephrasing the lemma into the language of Def.~\ref{single_shot_def}.

\section{Soundness and energy barriers}
\label{Sec_Energy}

Here we discuss the relationship between the concept of code soundness and energy barriers in physical systems, resulting in a proof of Thm.~\ref{THM_soundness_implies}.  The reader ought to ensure familiarity with the introductory material in subsections~\ref{Intro_soundness} and \ref{Intro_energy}.  Aharonov and Eldar remarked in Ref.~\cite{aharonov2015quantum} that codes with good soundness lead to large energy barriers, though they were interested in a strictly stronger definition of soundness.

A key lemma is the following
\begin{lem}
	\label{Claim_EnergyB}
	Consider a $[[n,k,d_Q]]$ quantum code with checks $\mathcal{M}$ that is $(t,f)$-sound and where all qubits are involved in no more than $C$ checks. It follows that the energy barrier is at least $f^{-1}(w)$ where $w=\mathrm{min}[ (t-1)/C , (d_Q-1)/2 ]$ and $f^{-1}$ is the inverse of the soundness function.
\end{lem}
For any walk of states $\{\psi_0,  \psi_1 , \psi_2, \ldots \psi_L \}$ we have a sequence of Pauli operators $\{  \id , E_1, E_2, \ldots E_L \}$, so that $\ket{\psi_j}=E_j\ket{\psi_0}$ and
$E_j E_{j-1}^{\dagger}=E_j E_{j-1}=P_j$ is a one qubit Pauli error (the local error condition). For every $E_j$ in the sequence we consider the reduced weight
\begin{equation}
	\wt_{\mathrm{red}} (E) := \mathrm{min}_V \{ \mathrm{wt}(E V ) :  V \in \mathcal{P} , \sigma(V)=0 \} ,
\end{equation}
where the minimisation is over all Pauli $V$ with trivial syndrome. Note that reduced weight is slightly difference from min-weight since the minimisation is over a bigger group than the code stabiliser. Herein we use $V_j$ to denote Pauli operators that achieve the above minimisation.  Since $\sigma(V_j)=0$ every $V_j$ is either a stabiliser or a nontrivial logical operator.  By the groundstates and orthogonality property, it follows that $V_0=\id$ and $V_L=E_L$.   So the sequence starts with a stabiliser and ends with a nontrivial logical operator.  Therefore, there must exist a $j^{\star}$ such that $V_{j^{\star}}$ is a stabiliser and $V_{j^{\star}+1}$ is a nontrivial logical operator.  Therefore, $V_{j^{\star}} V_{j^{\star}+1}$ must also be a nontrivial logical operator and so 
\begin{equation}
	\label{Eq_Fstar}
	d_Q \leq \wt( V_{j^{\star}}  V_{j^{\star}+1} ) .
\end{equation}  
Furthermore, we have
\begin{align} \nonumber
\wt( V_{j^{\star}}  V_{j^{\star}+1} ) = &  \wt( V_{j^{\star}}  V_{j^{\star}+1} E_{j^{\star}}  E_{j^{\star}}^\dagger E_{j^{\star}+1 } E_{j^{\star}+1 }^\dagger )   \\
= &  \wt( V_{j^{\star}} E_{j^{\star}}   V_{j^{\star}+1} E_{j^{\star}+1 }  E_{j^{\star}}  E_{j^{\star}+1 }) ,
\end{align}
and using the triangle inequality twice we have
\begin{align} \nonumber
	\wt( V_{j^{\star}}  V_{j^{\star}+1} ) \leq  &  \wt( V_{j^{\star}}  E_{j^{\star}} )	+ \wt(  V_{j^{\star}+1} E_{j^{\star}+1 } ) \\ \nonumber
 & + \wt( E_{j^{\star}} E_{j^{\star}+1} ) \\
= &  \wt_{\mathrm{red}}(E_{j^{\star}}) + \wt_{\mathrm{red}} (E_{j^{\star}+1}) + 1.
\end{align}	
We have used $\wt_{\mathrm{red}} (E_j)= \wt ( V_jE_j)$ on the first two terms and the local errors condition on the last term.  Combining this with Eq.~\eqref{Eq_Fstar}, leads to
\begin{align}
d_Q & \leq 	2 \mathrm{max}[\wt_{\mathrm{red}}(E_{j^{\star}}) ,\wt_{\mathrm{red}} (E_{j^{\star}+1}) ] + 1 ,
\end{align}	 
and so
\begin{align}
\frac{d_Q -1}{2}& \leq 	  \mathrm{max}[ \wt_{\mathrm{red}}(E_{j^{\star}}) , \wt_{\mathrm{red}}(E_{j^{\star}+1}) ] .
\end{align}	 
Consider the sequence of reduced weights $\{  \wt_{\mathrm{red}}(E_{0}) ,  \wt_{\mathrm{red}}(E_{1}) , \ldots , \wt_{\mathrm{red}}(E_{n}) \}$.  The sequence starts and ends with zero and at some point must reach $(d_Q-1)/2$ or higher.  Furthermore, the local error condition entails that $|\wt_{\mathrm{red}}(E_{j+1})  - \wt_{\mathrm{red}}(E_{j}) |$ is either 0 or 1 and so the sequence of reduced weights must include every integer from 0 to $(d_Q-1)/2$.  Therefore, we can set $w$ equal to $\min[t/C , (d_Q-1)/2]$ and there must exist an $E_{j}$ with $\wt_{\mathrm{red}}(E_{j})=w$.    Next, we consider the syndrome $\sigma (E_j)$ and note that $\sigma( E_j ) = \sigma (E_j V_j) $ where $\wt( E_j V_j )=\wt_{\mathrm{red}}(E_{j})$.  The LDPC condition of the code ensures that for any $E$ we have $|\sigma(E)| \leq C \wt(E)$.  Therefore, for the $E_j$ with $\wt_{\mathrm{red}}(E_{j})=w$ we have $|\sigma (E_j) | \leq C w$.  Since $w \leq (t-1)/C$ we have $|\sigma(E_j) | \leq t-1 < t$ and the soundness property can be deployed to conclude that $f^{-1}(w) \leq |\sigma(E_j)|$.  Since this holds for every possible walk, $f^{-1}(w)$ gives a lower on the energy barrier and we have proved Lem.~\ref{Claim_EnergyB}.

From Lem.~\ref{Claim_EnergyB} we can quickly obtain a proof of Thm.~\ref{THM_soundness_implies}.  We consider an infinite family of $[[n,k,d_Q]]$ codes with an LDPC check family $\mathcal{M}$ with good soundness.  That is, the codes are $(t_n,f)$-sound such that: the soundness function $f \in O(x^a)$ is independent of $n$; and $t_n$ grows as $\Omega(n^b)$ for some constants $a$ and $b$.  We further assume that the code distance $d_Q$ grows as $\Omega(n^c)$ for some constant $c$.   Since $d_Q \in \Omega(n^c)$ and $t \in \Omega(n^b)$, we can choose $w = \min[t/C , (d_Q-1)/2] \in \Omega(n^{\mathrm{min}[c,b]})$ in Lem.~\ref{Claim_EnergyB}.  It follows that the energy barrier scales as $ \Omega(n^{\mathrm{min}[c,b] / a})$  since $f \in O(x^a)$ and so $f^{-1} \in \Omega(x^{1/a})$.   Therefore this check family has a  macroscopic energy barrier.  Notice that soundness is not the only ingredient in the proof, the LDPC condition is also crucial.  It is unclear whether a similar result can be shown without the LDPC condition.  

We remark that the converse statement would be that any LDPC check family with a macroscopic energy barrier has good soundness.  We have neither proof nor counterexample and so the status of this converse statement remains open.

Bravyi and Terhal proved that no 2D topological stabiliser codes have a macroscopic energy barrier~\cite{bravyi2009no}.   Therefore, such codes cannot have good soundness as we stated in corollary~\ref{No_go_corollary}.  This is nearly a statement that single-shot error correction is impossible in 2D topological stabiliser codes and we believe this to be the case. Though one must be cautious as we have shown good soundness to be a sufficient condition for single-shot error correction but not a necessary one.  Clearly, if a code does not have good soundness then minimum weight decoding (in the sense of Def.~\ref{Def_SS_decode}) can lead to large weight residual error.  However, if one deviates from the minimum weight decoding strategy then the picture becomes less clear.  For instance, one strategy might be that when the minimum weight solution is high weight, we do not attempt to return the system to the codespace but instead apply a partial recovery.  For instance, if we observe two far apart checks with ``-1" outcomes in the 2D toric code, then we could apply a partial recovery that reduces the distance between these checks.  Indeed, there are cellular automata decoders for the 2D toric code that behave just like this~\cite{Har04,herold2015cellular,breuckmann2016local,herold2017cellular}.  These fail to qualify as single-shot decoders in the usual sense as they rely on the syndrome history (partially stored in a cellular automata).  But they highlight that single-shot error correction might be possible using an imaginative decoder approach based on partial recoveries.  

\section{Good soundness for all codes}
\label{Sec_Simple}

It is common to conflate a quantum error correction code with a set of checks $\mathcal{M}$ that generate the stabiliser.  But there are many choices of checks for any given code.  Crucially, the soundness properties depend on the set of checks.  Here we prove Thm.~\ref{THM_soundness_simple}, which roughly states that for any code we can find a check set with good soundness properties.   The proof follows from the following lemma.
\begin{lem}
	\label{Lem_PureErrors}
	Given an $[[n,k,d_Q]]$ quantum error correction code with stabiliser $\mathcal{S}$ there exists a minimal set of generators $\mathcal{M}=\{ M_1, M_2,  \ldots ,  M_{n-k}  \}$ and associated Pauli errors $\mathcal{E}=\{ E_1, E_2,  \ldots ,  E_{n-k}  \}$ such that: (1) $[ M_i , E_j] \neq 0$ if and only if $i=j$; and (2) every $E_j$ acts non-trivially on only a single qubit and so $\wt(E_j)=1$.
\end{lem}	
We first consider the consequence of this lemma. Given such a set of checks, it follows that if $s$ is a syndrome unit vector (so $|s|=1$) with a 1 entry in the $j^{\mathrm{th}}$ location, then $s=\sigma(E_j)$ (recall Eq.~\eqref{Eq_syndrome_def}).  More generally, $s$ can be written as a sum of $|s|$ unit vectors and therefore $s=\sigma(E)$ where 
\begin{equation}
	E = \prod_{j : s_j =1} E_j.
\end{equation}
Since $\wt(E_j)=1$ we have $\wt(E) \leq |s|$ (with more work one can prove equality).  Therefore, the checks are $(t,f)$-sound with $t=\infty$ and $f(x)=x$ since:   the argument holds for any weight syndrome, and so the value of $t$ is unbounded;  and the weight of the physical error is no more than the weight of the syndrome, so we have $f(x)=x$. 

The proof of Lem.~\ref{Lem_PureErrors} is essentially a step in the proof  Lem.~2 of Ref.~\cite{Camp10a}.   In Ref.~\cite{Camp10a}, it is shown that upto to qubit labelling and local Clifford unitaries, the generators $M_j$ can be brought into a diagonalised form inspired by the graph state formalism.  In this form, $M_j$ acts on the $j^{\mathrm{th}}$ qubit with Pauli $X$.  On all others qubits with labels 1 through to $n-k$, the operator $M_j$ acts as either Pauli $Z$ or the identity.  Therefore, Pauli $Z$ acting on qubit $j$ anticommutes with generator $M_j$ and commutes with all other generators.    Accounting for local Cliffords and original qubit labelling, the required $E_j$ may act on a different qubit and may be different from Pauli $Z$, but it will be a single qubit Pauli.  This completes the proof.

The soundness properties proven above are extremely strong.  This leads to the counter-intuitive result that single-shot error correction is possible for any code and without any check redundancy.  The price to pay is that one must use a certain set of checks such as the diagonalised form above.  As such, if the checks are initially low weight (e.g. part of an LDPC check family) then this property may be lost as the diagonalisation process leads to high weight checks.  Indeed, we can prove the following strong limitation on diagonalisation methods.

\begin{claim}
	Consider a family of codes with checks in the diagonalised form used in the proof of Lem.~\ref{Lem_PureErrors}.  Assume also the diagonalised check family is LDPC, such that in every code no qubit is acted on by more than $C$ checks.  It follows that the distance is bounded $d_Q \leq C+1$ for all codes in the family.
\end{claim}

We prove this by constructing an explicit error $F$ that is not in the code stabiliser but $\sigma(F)=0$ and $\wt(F) \leq C+1$.  First, we let $P$ be some single qubit Pauli ($\wt(P) = 1$) acting on a qubit with label exceeding $n-k$.  By the LDPC property $|\sigma(P)| \leq C$.  Furthermore, following previous arguments there exists an $E$ acting on the first $n-k$ qubits such that $\sigma(E)=\sigma(P)$ and $\wt(E) \leq |\sigma(P)|$.     Combined  $\wt(E) \leq |\sigma(P)|$ and $|\sigma(P)| \leq C$ entail $\wt(E) \leq C$.  Setting $F=EP$, we have that 
\begin{equation}
	\sigma(F)=\sigma(E)+\sigma(P)=2 \sigma(E) = 0
\end{equation}
and 
\begin{equation}
	\wt(F) \leq \wt(E)+\wt(P) \leq C + 1.
\end{equation}
Lastly, we need to show that $F$ is not an element of the stabiliser.  First we note that $F \neq \id$ since $E$ and $P$ act on disjoint sets of qubits.  Next, let us assume to the contrary that $F$ is a non-trivial element of the stabiliser. Then there is some non-empty set $J \subseteq \{ 1, \ldots, n-k \}$ such that 
\begin{equation}
	F = \prod_{j \in J} M_j.
\end{equation}	
 Following the argument in the proof of Lem.~\ref{Lem_PureErrors}, let us assume that each $M_j$ acts with Pauli $X$ on the $j^{\mathrm{th}}$ qubit.  But all $M_{k \neq j}$ act on the $j^{\mathrm{th}}$ qubit with either Pauli $Z$ or the identity.  Therefore, for every $j \in J$ we have that $F$ acts on the $j^{\mathrm{th}}$ qubit with either $X$ or $Y$.  Since $J$ is non-empty there is at least one qubit with index between 1 and $n-k$ such that $F$ acts as either $X$ or $Y$.  However, $F=EP$ where $E$ acts on the first $n-k$ qubits with either $Z$ or $\id$.  Since $P$ acts on one of the last $k$ qubits, we see that $F$ can not be a stabiliser and must instead be a non-trivial logical operator.
 
The LDPC property is highly desirable and so too is growing code distance.   Therefore, we need an alternative route to good soundness.

\section{Tanner graphs, chain complexes and homology theory}
\label{Sec_Homology_theory}

From here on we specialise to codes with checks  $\mathcal{M}$ that can be partitioned into checks in the $Z$ and $X$ Pauli basis. For such codes, we describe quantum codes in a graphical language that extends on the classical use of Tanner graphs.  We will explain the correspondence between the graphical representation and a linear algebra description in terms of concepts from algebraic topology.  

\begin{figure*}
	\includegraphics{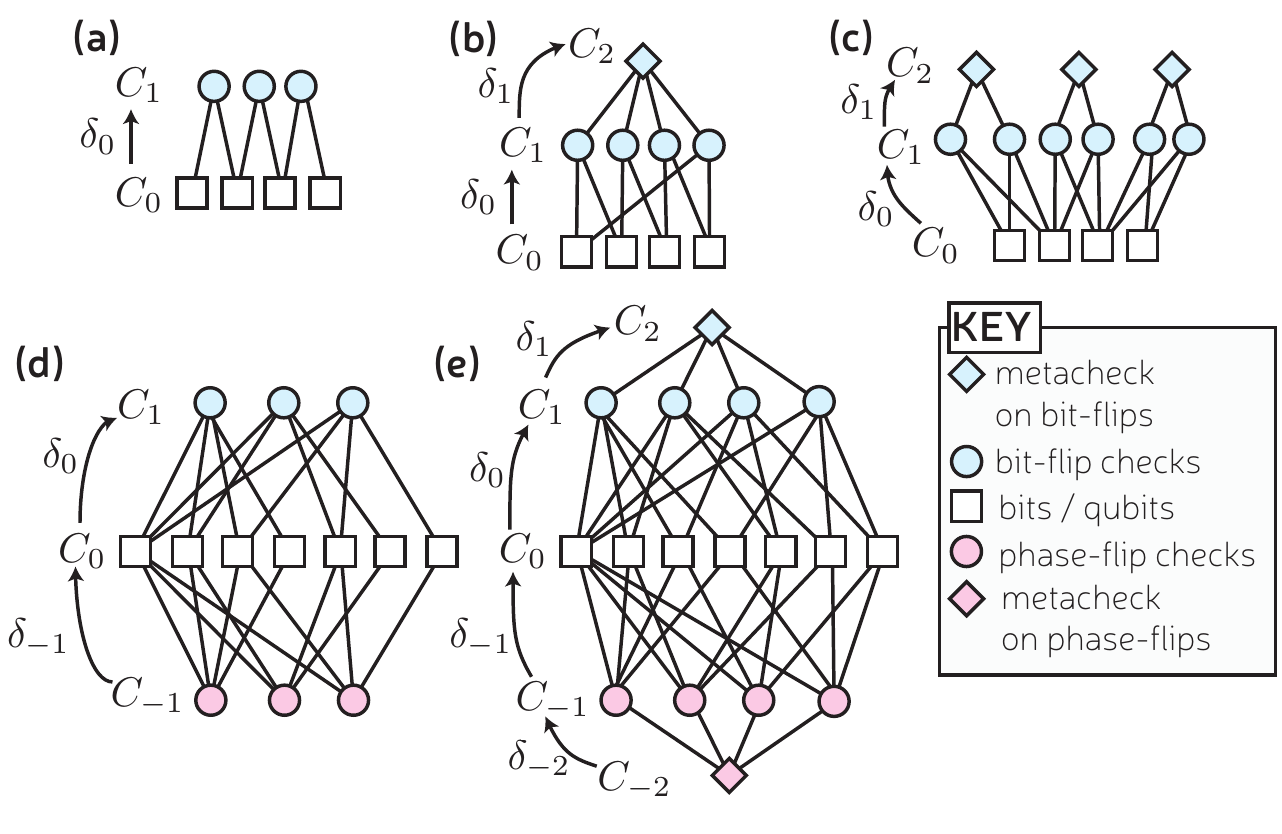}
	\caption{A graphical representation of some example classical and quantum error correcting codes, including scheme for parity check measurements and metachecks.  (a) the 4 bit classical repetition code; (b) the 4 bit classical repetition code with an additional check and corresponding metachecks;  (c) the 4 bit classical repetition code with repeated checks and corresponding metachecks; (d) the 7-qubit Steane code; (e)  the 7-qubit Steane code with additional checks and corresponding metachecks.  The symbol $\delta_{j}$ is a matrix describing the connectivity between vertices in set $C_j$ and $C_{j+1}$.  It can also be considered as a linear map known as the boundary map in homology theory.}
	\label{Fig_Hasse}
\end{figure*}

Several example graphs are given in Fig.~\ref{Fig_Hasse}.  In every case, the graph breaks up into $D+1$-partitions and we will refer to $D$ as the length of the graph. Each partition comes with a set of vertices $C_j$.   We use a binary matrix $\delta_j$ to describe the adjacency between vertices in $C_j$ and $C_{j+1}$.  Specifically, matrix $\delta_j$ has a ``1" in entry $(a,b)$ if and only if the $b^{\mathrm{th}}$ vertex in $C_j$ is connected to the $a^{\mathrm{th}}$ vertex in $C_{j+1}$.  Therefore, $\delta_0$ is the well-known parity check matrix of a classical code.  Furthermore, $\delta_0$ is the  parity check matrix for bit-flip ($X$) errors in a quantum code.  Using superscript $T$ for transpose, the matrix $\delta_{-1}^T$ is the  parity check matrix for phase-flip ($Z$) errors in a quantum code.   

We conflate thinking of $C_j$ as a set of vertices and also as a binary vector space $\mathbb{Z}_2^{n_j}$ where $n_j$ denotes the number of vertices in $C_j$.  A unit vector $\hat{u}$ has only a single entry with value 1 and identifies single vertex in $C_j$.  Therefore, given a pair of unit vectors $\hat{u} \in C_j$ and $\hat{v} \in C_{j+1}$, we have $\hat{v}^T \delta_j \hat{u}=1$ if and only if the corresponding vertices are connected.   Therefore, given a unit vector $\hat{u} \in C_1$ identifying a measurement (or check) for bit-flip errors, the vector $\delta_0^T \hat{u} $ identifies the (qu)bits involved in that  check.  We use the notation
\begin{align}
X[u] & :=\otimes_j X_j^{u_j} , \\
Z[v] & :=\otimes_j Z_j^{v_j} , 
\end{align}
where $u$ and $v$ are binary vectors.  The graph should be read as not just defining a code but also the measurement scheme. So for every unit vector $\hat{u}$ in $C_1$, the graphical formalism stipulates that we measure the operator $Z[\delta_0^T \hat{u}]$.  So in our earlier notation $Z[\delta_0^T \hat{u}]$ would be a member of $\mathcal{M}$ and is a stabiliser of the code.  Since the stabiliser is a group, we have  that $Z[\delta_0^T u]$ is a stabiliser for any vector $u \in C_1$.   Similarly, $X[\delta_{-1} v]$ is a stabiliser of the code for every  $v \in C_{-1}$.    Operators $X[\delta_{-1} v]$ and $Z[\delta_0^T u]$ will commute if and only if $(\delta_0^T u)^T\delta_{-1} v=u^T \delta_0 \delta_{-1} v = 0$ where all such equations should be read using addition modulo 2.  Since we need all such operators to commute, we require that $\delta_0 \delta_{-1}=0$.   Conversely, if $X[e]$ with $e \in C_0$ is an error, the vector $\delta_0e$ is the $Z$-measurement syndrome assuming ideal measurements.

\begin{figure*}
	\includegraphics{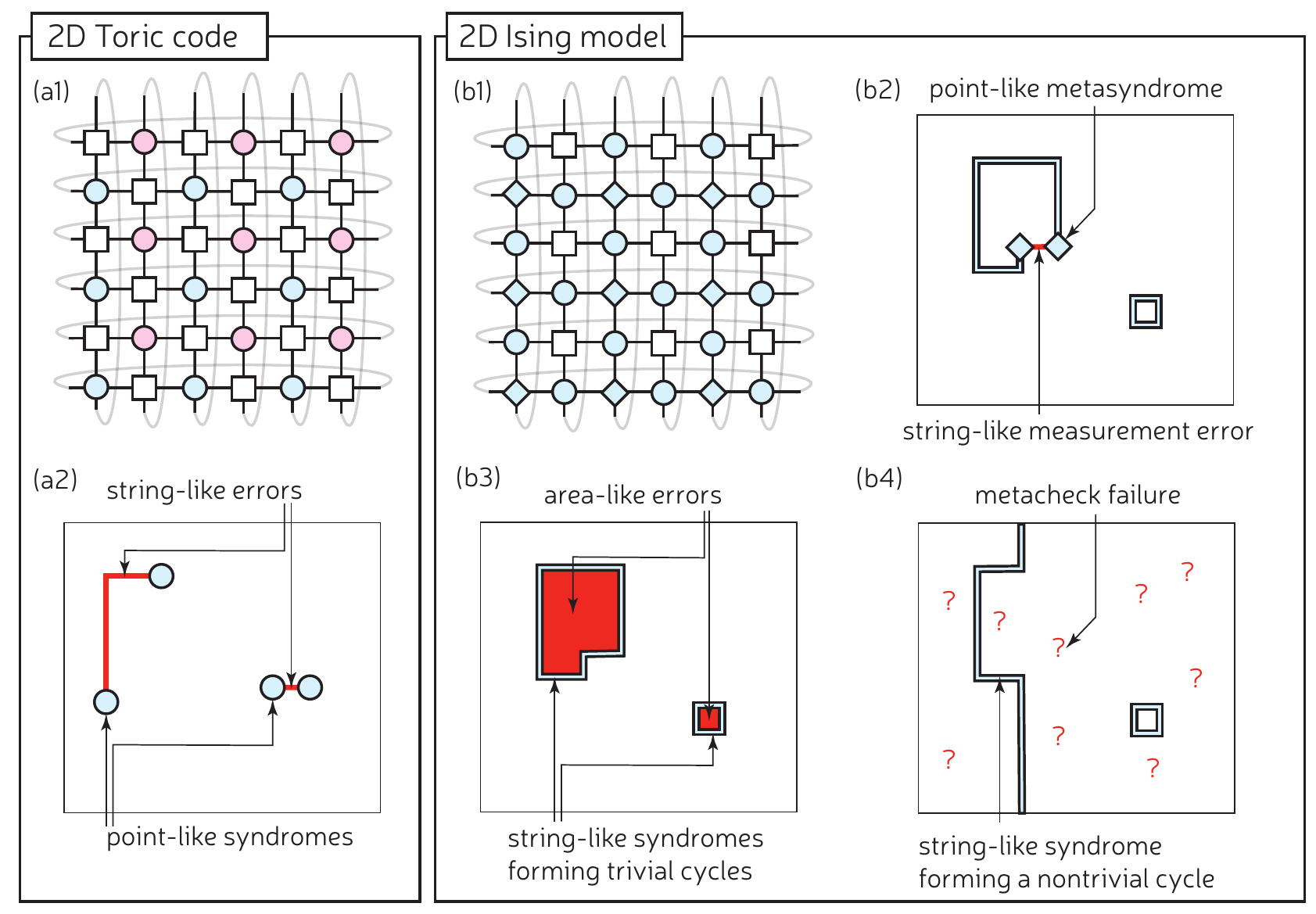}
	\caption{In (a) we illustrate the 2D toric code.  Part (a1) describes the toric code using the vertex labelling from Fig.~\ref{Fig_Hasse} with grey curved lines highlighting the periodic boundary conditions of the torus. Part (a2) shows the relationship between error and syndromes.  Notice that a weight 2 syndrome (two endpoints) could require an arbitrarily long string to produce the syndrome.  Therefore, the code does not have good soundness.  In (b) we illustrate the 2D Ising model as a classical error correction code.  Part (b1) again uses the vertex labelling from Fig.~\ref{Fig_Hasse}.  Notice that (b1) represents the same graph as (a1) but with the different types of vertex changing role.  Part (b2) shows a measurement error that is detected by  metachecks.  Part (b3) shows a measurement syndrome that passes all metachecks (i.e. it would be the corrected syndrome of (b2)).  The red region shows an error pattern that generates the syndrome.  Notice that the size of the physical error scales at most quadratically with the size of the syndrome. Therefore, the code does have good soundness.  Part (b4) show a metacheck failure.  There is a syndrome that spans the code and forms a non-trivial cycle.  Due to periodic boundary conditions there is no error region with this syndrome as its boundary.}
	\label{Fig_ToricIsing}
\end{figure*}

In homology theory, this whole structure is called a chain complex and the operators $\delta_j$ are called boundary maps provided the relation $\delta_{j+1} \delta_j = 0$ holds for all $j$.  Therefore, given a homological chain complex the commutation relations are automatically satisfied since $\delta_0 \delta_{-1}=0$.  Remarkably, requiring  $\delta_{j+1} \delta_j = 0$ not only gives us the required commutation relations but also ensures that the metachecks are suitably defined.  We will show this formally.  Consider a physical error $X[e]$.   It will generate $Z$-syndrome $\delta_0e$ assuming no measurement errors.  Since there are no measurement errors, the metasyndrome $x=\delta_1 \delta_0 e$ ought to be the all zero vector, which is ensured if $\delta_1 \delta_0 =0$.  

Let us connect this back to the notation used in the first part of this paper.  The check set is
\begin{equation}
\mathcal{M}=( Z[\delta_0^T  \hat{u}_1], \ldots ,  Z[ \delta_0^T \hat{u}_{n_{1}}] , X[\delta_{-1} \hat{v}_1], \ldots , X[\delta_{-1} \hat{v}_{n_{-1}}]  )
\end{equation} 
where $\hat{u}_j$ and $\hat{v}_j$ are unit vectors with the unit in the $j^{\mathrm{th}}$ location.  Any Pauli error can be expressed as $E = X[e]Z[f]$ for some vectors $e$ and $f$.  The syndrome of this Pauli is then the combination of the $Z$ and $X$ syndromes, so that
\begin{equation}
   \sigma( X[e]Z[f] ) = \left( \begin{array}{c}
    \delta_0e \\
    \delta_{-1}^T  f 
   \end{array}  \right) .
\end{equation}
Furthermore, the whole metasyndrome matrix has block matrix form
\begin{equation}
H = \left( \begin{array}{cc}
\delta_1 & 0 \\
0 & \delta_{-2}^T  
\end{array}  \right) .
\end{equation}
From this we see that the condition required earlier (that $H \sigma(E)=0$ for all Pauli $E$) follows from the fundamental property of chain complexes, specifically $ \delta_1 \delta_0 = 0$ and $ \delta^T_{-2} \delta^T_{-1} = 0$.

Next, we study some parameters of chain complexes.  We use $n_j$ to denote the number of vertices in $C_j$, and equivalently the dimension of the associated vector space $\mathbb{Z}_2^{n_j}$.  The matrix $\delta_j$ will have $n_j$ columns and $n_{j+1}$ rows.  An important parameter is the $j^{\mathrm{th}}$ Betti number, which we denote $k_j$.   For our purposes, it suffices to define
\begin{equation}
\label{Eq_Betti}
k_j :=   \mathrm{nullity}(\delta_j) -  \mathrm{rank}(\delta_{j-1}) .
\end{equation}	
Here, $\mathrm{nullity}$ is the dimension of the kernel, denoted $\ker(\delta_j)$, which is the space of vectors $u$ such that $\delta_j u=0$.  The $\mathrm{rank}$ is the number of linearly independent rows in a matrix. Alternatively, the $\mathrm{rank}$ is equal to the dimension of the image, denoted $\im(\delta_{j-1})$, which is the space of vectors $v$ such that there exists a $u$ satisfying $v=\delta_{j-1}u$.  Those familiar with homology theory may prefer to think of $k_j$ as the dimension of the $j^{\mathrm{th}}$ homology group $\mathcal{H}_j= \mathrm{ker}(\delta_j) / \mathrm{im}(\delta_{j-1})$.   This counts the number of different homology classes at a particular level of the chain complex.  Let $c$ be an element of $C_j$.  If $c \in \mathrm{ker}(\delta_j)$ then we say $c$ is a cycle.  However, for any $c \in \mathrm{im}(\delta_{j-1})$ it immediately follows from $\delta_j \delta_{j-1}=0$ that also $c \in \mathrm{ker}(\delta_j)$ and such a cycle is said to be a trivial cycle.  On the other hand, if $c \in \mathrm{ker}(\delta_j)$ but  $c \notin \mathrm{im}(\delta_{j-1})$ then $c$ is a non-trivial cycle.  If any non-trivial cycles exist then $k_j>0$, and the value of $k_j$ counts the number of different non-trivial cycles (factoring out homological equivalence).   Note that for $k_j$ with the lowest value of $j$ in the chain complex, the matrix $\delta_{j-1}$ is not defined and so Eq.~\eqref{Eq_Betti} should be read with $\delta_{j-1}$ substituted by the zero matrix.  Similarly, for the largest possible $j$ value we must take $\delta_j$ as the zero matrix.

One can similarly look at the cohomologies   
\begin{equation}
\label{Eq_CoBetti}
k_j^T :=  \mathrm{nullity}(\delta^T_{j-1}) -  \mathrm{rank}(\delta^T_{j}) .
\end{equation}
Poincar\'{e} duality entails that $k_j^T=k_j$ and for completeness we give a simple proof in  App.~\ref{App_Poincare} using only linear algebra.  For quantum codes, $k_0$ is important as it gives the number of logical qubits encoded by the code.  It is useful for us to also to consider $k_j$ for other values of $j$.  For instance, in a code with metachecks, $k_1$ is the number of classes of syndromes $x$ such that they pass all the metachecks ($\delta_1 x =0 $) but there does not exist an explanation in terms of qubit errors ($\nexists e$ such that $x=\delta_0 e $).  

In the context of error correction, we are interested not just in the number of non-trivial cycles, but also their minimum distance.  As such, we define
\begin{align} \label{Eq_distances}
d_j & := \mathrm{min} \{ |c|  :  c \in \mathrm{ker}(\delta_j) , c \notin \mathrm{im}(\delta_{j-1})    \} , \\ \nonumber
d_j^T & := \mathrm{min} \{ |c|  :  c \in \mathrm{ker}(\delta^T_j) , c \notin \mathrm{im}(\delta^T_{j+1})    \} ,
\end{align}
where $|c|:=\sum_j c_j$ is the Hamming weight.  We use the convention that $d_j=\infty $ whenever $k_j=0$ and similarly  $d_j^T=\infty$ whenever $k_{j+1}^T$=0.   We know of no simple relationship between $d_j$ and $d_j^T$.  This is enough for us to define the usual parameters of the corresponding $[[n,k,d_Q]]$ quantum code as $n=n_0$, $k=k_0$ and $d_Q=\mathrm{min}[d_0, d_{-1}^T]$.  However, we also introduce a new parameter that we call the single-shot distance as follows.
\begin{defin}[Single-shot distance]
	Given a length-4 chain complex we define the single-shot distance as $d_{ss}:=\mathrm{min}[d_1, d_{-2}^T]$ where $d_1$ and $d_{-2}^T$ are special cases of Eq.~\eqref{Eq_distances}.
\end{defin}
The single-shot distance relates to how many measurement errors can be tolerated before a failure occurs that we call a metacheck failure.  In a metacheck failure, the syndrome has no explanation in terms of qubit errors.  See Fig.~\ref{Fig_ToricIsing}b4 for an example of metacheck failure in the 2D Ising model with periodic boundary conditions. 

Let us review different ways we can use this formalism. Consider a length-1 chain complex $C_0 \rightarrow_{\delta_0} C_1$.  We can consider the vertices in the zeroth level as bits and the first level as parity checks.  Thus a length-1 chain complex can be regarded as a classical code. Consider a length-2 chain complex $C_{-1} \rightarrow_{\delta_{-1}} C_0 \rightarrow_{\delta_0} C_1$.   This could represent either a quantum code (without any metachecks) or alternatively a classical code equipped with metachecks.  In the classical case, our convention is to increment all the indices by one to have $C_{0} \rightarrow_{\delta_{0}} C_1 \rightarrow_{\delta_1} C_2$.  We choose this convention such that $C_{0}$ always labels the physical bits or qubits.   In Fig.~\ref{Fig_ToricIsing}a1 and Fig.~\ref{Fig_ToricIsing}b1 we show two graphs representing length-2 chain complexes.  The graphs are identical except in Fig.~\ref{Fig_ToricIsing}a1 it represents a quantum code and in Fig.~\ref{Fig_ToricIsing}b1 it represents a classical code with metachecks.

Given a length-4 chain complex, the additional layers of homology describe metachecks on the $X$ and $Z$ checks.  Note that the additional layers of the chain complex have no direct effect on the code parameters.  

We could also consider length-3 chain complexes with metachecks on either $X$ and $Z$ checks.  It is also plausible that a length-3 chain complex could support single-shot error correction of both error types by using a form of gauge fixing such as proposed in 3D colour codes~\cite{bombin2015single}. However, we will not explore this here.

We also need to translate the notion of soundness into the language of chain complexes
\begin{defin}[Soundness of maps] \label{ssDEF}
	Let $t$ be an integer and $f : \mathbb{Z} \rightarrow \mathbb{R}$ be some function called the soundness function.  Given a linear map $\delta$, we say it is $( t, f)$-sound if for all $r$ such that $| \delta r| < t$, it follows that:
	\begin{align}
	x =	| \delta r|   & \implies \mathrm{min} \{ | r' | :   \delta r' = \delta r  \}  \leq f(x) .
	\end{align}
	Furthermore, we say a quantum error correcting code  is $( t, f)$-sound if the above holds for both $\delta_0$ and $\delta_{-1}^T$.     For a classical error correcting code this is required for just $\delta_0$.
\end{defin}
This is less general than the earlier Def.~\ref{RoughssDEF} since the above only applies to CCS codes whereas our earlier definition was valid for any stabiliser code.  However, it should be clear that any CCS code satisfying Def.~\ref{ssDEF} will also satisfy Def.~\ref{RoughssDEF}.  We saw earlier that 2D topological codes cannot have good soundness and we illustrate this in Fig.~\ref{Fig_ToricIsing}a.  Whereas, for the 4D toric code, with an appropriate choice of checks,  geometric arguments show that low weight syndromes can always be generated by small weight errors.  To visualise this, it is easier to instead think of the 2D Ising model as a classical error correcting code.  In such a code, syndrome cycles have a weight equal to their perimeter and the error generating the syndrome has weight equal to the area  (see Fig.~\ref{Fig_ToricIsing}b3).  The area of a 2D region can be no more than $x^2/8$ of the perimeter length $x$ and so the Ising model has a quadratic soundness function.  Therefore, it can be helpful to think of soundness as describing the geometric area law relationship between syndromes and errors, albeit in purely algebraic terms. 

Check redundancy provides consistency conditions that one can inspect for evidence of measurement errors.   These checks on checks are illustrated in Fig.~\ref{Fig_Hasse} using diamonds. We call these metachecks.  They do not represent a physical measurement but classical postprocessing on the measurement outcomes.  That is, for a given metacheck node we calculate the parity of all the checks it is connected to. If this parity is odd, a measurement error must have occurred on one of the adjacent nodes.   Recall that we quantify the amount of redundancy in a measurement scheme as the ratio between the number of measurements performed (which equals $n_1 + n_{-1}$) and the minimum number required to generate the stabiliser of the code (which equals $n_0-k_0$). We use $\upsilon$ to denote this ratio, so that
\begin{equation}
\label{Eq_Redundancy_Def}
\upsilon = \frac{n_1 + n_{-1}}{n_0-k_0},
\end{equation}
with $\upsilon=1$ indicating no redundancy.   In Fig.~\ref{Fig_Hasse} we give examples of codes with such redundancy  (Fig.~\ref{Fig_Hasse}b, Fig.~\ref{Fig_Hasse}c and  Fig.~\ref{Fig_Hasse}c).  We are interested in  check families where $\upsilon$ is no more than a small constant factor. 

\section{Constructing single-shot codes}
\label{Sec_Constructions}

Here we show how the homological product can be used to construct new codes supporting single-shot error correction.  This will culminate in a proof of Thm.~\ref{THM_construct} though the techniques allow for a broader range of constructions, including codes where the single-shot distance is finite.

\subsection{A single application constructions}
\label{Sec_firstHomo}

As a warm-up, we begin by considering a single application of the homological product.  Our approach is to take a length-1 chain complex (e.g. a conventional classical code) and use the homological, or hypergraph, product to build a length-2 chain complex with the desired properties.  In general, one could take two different input classical codes and combine them together using these techniques, but for simplicity we take both input codes to be the same.  Furthermore, there are a few different notions of the homological product.  For instance, Bravyi and Hastings use a simplified variant that they call the single sector homological product, whereas we will use a more standard textbook variant that Bravyi and Hastings would call a multi sector homological product~\cite{bravyi2014homological}.  Furthermore, there is some freedom in the notation and we use a convention such that the homological product in this section is manifestly equivalent to the hypergraph product of Tillich and Zemor~\cite{tillich2014quantum}.

\begin{figure*}
	\includegraphics{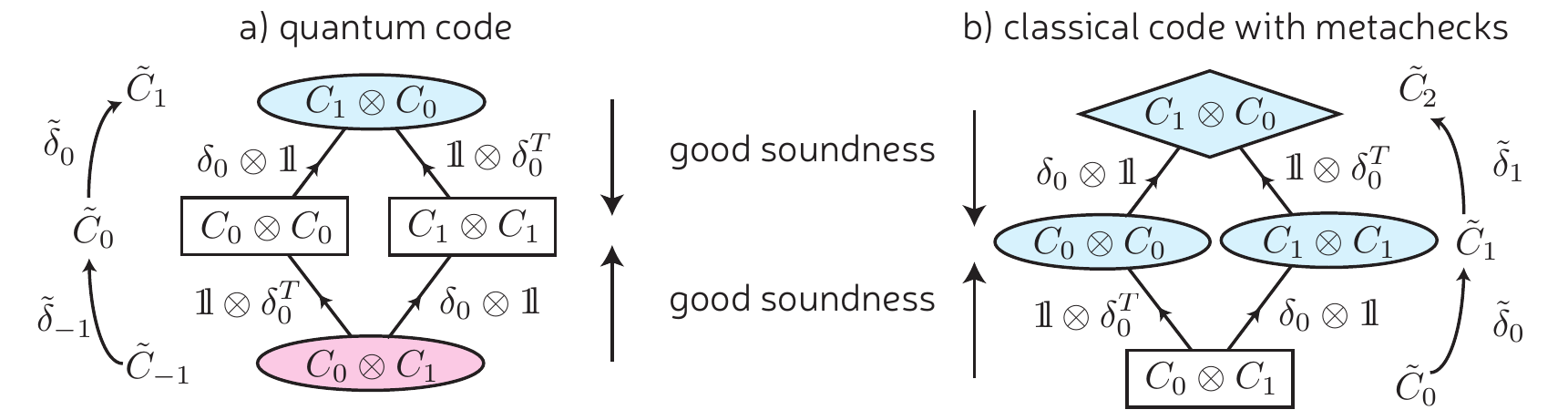}
	\caption{An overview of a single application of the homological product to generate a length-2 chain complex from a length-1 chain complex (that can be viewed as a classical code).  In (a) we label the chain-complex under the assumption that it defines a quantum code, and where the subscripts are consistent with the main text. In (a) we label the chain-complex under the assumption that it defines a classical code.  In order that $\tilde{C}_0$ denotes the bits, we have increments all the new subscripts by 1.  Throughout we use rectangles to show a collection of bit/qubit vertices; we use ovals to show a collection of checks; and diamonds to show a collect of metachecks.}
	\label{Fig_SingleHP}
\end{figure*}

Given a chain complex $C_0 \rightarrow_{\delta_0} C_1$ we can define a new chain complex $  \tilde{C}_{-1} \rightarrow_{\tilde{\delta}_{-1}} \tilde{C}_{0} \rightarrow_{\tilde{\delta}_0} \tilde{C_{1}} $ of the form
\begin{equation}
	  C_{0} \otimes C_{1} \rightarrow_{\tilde{\delta}_{-1}} (C_{0} \otimes C_{0}) \oplus  (C_{1} \otimes C_{1})  \rightarrow_{\tilde{\delta}_0} C_{1} \otimes C_{0}. 
\end{equation}
The notation $\otimes$ represents the tensor product.  For example, if $a \in C_{0}$ and $b \in C_{1}$ then $a \otimes b \in C_{0} \otimes C_{1} $, and the space $C_{0} \otimes C_{1} $ further contains any linear combinations of such vectors.  The symbol $ \oplus$ represents a direct product.  For instance, vectors in $(C_{0} \otimes C_{0}) \oplus  (C_{1} \otimes C_{1})$ can be written as $w =u \oplus v$ where $u \in (C_{0} \otimes C_{0})$ and $v \in (C_{1} \otimes C_{1})$.   All vectors should be read as column vectors and so the direct product of vectors can also be read as stacking these vectors
\begin{equation}
u \oplus v = \left(  \begin{array}{c}
u  \\
v \end{array}
\right) .
\end{equation} 
We will use the weight identities $| u \otimes v| = |u| \cdot |v|$ and  $| u \oplus v| = |u| + |v|$.  The boundary map $\tilde{\delta}_{-1}$ is defined such that for product vectors $a \otimes b \in   C_{0} \otimes C_{1} $, we have
\begin{equation}
\label{BoundMap}
	\tilde{\delta}_{-1} ( a \otimes b ) =  (a \otimes (\delta_0^T b) )  \oplus ((\delta_0a) \otimes b ),
\end{equation}
and it extends linearly to non-product vectors.  This is often more concisely denoted as $\tilde{\delta}_{-1} = (\id \otimes \delta_0^T) \oplus (\delta_0 \otimes \id)$.  The boundary map $\tilde{\delta}_0$ is defined such that for product vectors $a \otimes b \in   C_{0} \otimes C_{0} $ and $c \otimes d \in   C_{1} \otimes C_{1} $, we have
\begin{equation}
\tilde{\delta}_0 ( (a \otimes b)  \oplus (c \otimes d)  ) = (( \delta_0 a) \otimes b )+  (c \otimes (\delta_0^T d)  ) ,
\end{equation}
and again extending linearly to non-product vectors.  Both the new boundary maps can also be represented in block matrix form
\begin{align}
	\tilde{\delta}_{-1} & = \left(  \begin{array}{c}
		\id \otimes \delta_0^T \\
		\delta_0 \otimes \id
		\end{array} \right) , \\ \nonumber
\tilde{\delta}_0 & = \left(  \begin{array}{cc}
		\delta_0 \otimes \id & \id \otimes \delta_0^T 		
	\end{array} \right) .
\end{align}
From here it is easy to verify that they satisfy the requirement that 	$ \tilde{\delta}_0 	\tilde{\delta}_{-1} =  2( \delta_0 \otimes \delta_0^T) = 0$, where we have used that all mathematics is being performed modulo 2.  These matrices fully characterise the new chain complex and from them we can find graphs of the sort shown in Fig.~\ref{Fig_Hasse}.  We give a graphical overview in Fig.~\ref{Fig_SingleHP}.

Now we discuss the parameters of this new structure, with some of these results obtained in Ref.~\cite{tillich2014quantum}. Simple dimension counting tells us that the new chain complex has
\begin{align}
\label{Ntidle}
	\tilde{n}_{-1} & = n_0 n_1 , \\ \nonumber
	\tilde{n}_0 & = n_0^2  + n_1^2 , \\ \nonumber
	\tilde{n}_1 &=  n_0 n_1 .
\end{align}	
The dimension of the homological classes is more involved, but a well known result from homology theory (the K\"{u}nneth formula~\cite{hatcher2002algebraic,bravyi2014homological}) tells us that 
\begin{align}
\label{Ktidle}
	\tilde{k}_{-1}& = k_0 k_1 , \\ \nonumber
	\tilde{k}_0 & = k_0^2  + k_1^2 , \\ \nonumber
	\tilde{k}_1 &=  k_1 k_0 .
\end{align}	
The distance of the code is trickier yet again to prove and is not a standard quantity in homology theory.  Nevertheless, one can show that
\begin{align}    \label{Eq_distances1}
	\tilde{d}_{-1}  	& = d_0 d_0^T ,  \\    \label{Eq_distances2}
	\tilde{d}_0^T   & = d_0 d_0^T , \\ \label{Eq_distances3}
	\tilde{d}_0 &  \geq \mathrm{min} ( d_0 , d_0^T) ,  \\   \label{Eq_distances4}
	\tilde{d}_{-1}^T &  \geq \mathrm{min} ( d_0 , d_0^T)  .
\end{align}
We provide proofs in App.~\ref{App_distances} for Eq.~\eqref{Eq_distances1} and Eq.~\eqref{Eq_distances2}.  The results of Eq.~\eqref{Eq_distances3} and Eq.~\eqref{Eq_distances4} were shown by Tillich and Zemor~\cite{tillich2014quantum} but we give an independent proof in the homological formalism in App.~\ref{App_distances}.

Here we instead focus on the following lemma
	\begin{lem}[First soundness lemma]
		\label{Lem_Coexpand}
			Let $C_0 \rightarrow_{\delta_0} C_1$ be a chain complex.  Applying the above homological product we obtain a new chain complex where the map $\tilde{\delta}_{0}^T$ is $(t ,f)$-sound  and $\tilde{\delta}_{-1}$ is $(t ,f)$-sound with $f(x)=x^2/4$ and $t=\mathrm{min}(d_0, d_0^T)$.
	\end{lem}
We make no assumptions about the soundness properties of the original chain complex but find this emerges due to the nature of the homological product. However, if one knows that the original chain complex is sound, one could prove a stronger soundness result (with $f$ growing slower than $x^2/4$) for the new chain complex.   We prove this lemma in App.~\ref{App_Coexpand1} and next discuss its implications.

Using the above homological product, we can construct a quantum code with parameters $[[ \tilde{n}_0, \tilde{k}_0, d_Q ]]$ where $d_Q=\mathrm{min}[\tilde{d}_0^T, \tilde{d}_0]$.  These codes will not necessarily support single-shot error correction because the soundness property in Lem.~\ref{Lem_Coexpand} is not the property required by Thm.~\ref{THM_main}, which requires that $\tilde{\delta}_{0}$ and $\tilde{\delta}_{-1}^T$ have good soundness properties.  

Why prove Lem.~\ref{Lem_Coexpand} if it is does not directly provide quantum codes with single-shot capabilities?  First, in the next section we will make a second application of the homological product and Lem.~\ref{Lem_Coexpand} will be used, and so it is a stepping stone result.  Second, Lem.~\ref{Lem_Coexpand} is highly instructive as it gives a way to construct classical codes that exhibit single-shot error correction.  Let us explore this second point further.  A classical code with metachecks needs three layers of structure (recall Fig.~\ref{Fig_Hasse}) and our convention is that the subscript $0$ in $C_0$ always denotes the bits or qubits.   So for a classical code with metachecks, we want a chain complex of the form $  \tilde{C_{0}} \rightarrow_{\tilde{\delta}_{0}} \tilde{C_{1}} \rightarrow_{\tilde{\delta}_1} \tilde{C_{2}} $.  We can use the chain complex generated by the homological product by simply increasing all the subscripts by 1.  With these incremented subscripts, Lem.~\ref{Lem_Coexpand} tells us that  $\tilde{\delta}_{0}$ is $(d_0^T ,f)$-sound with $f(x)=x^2/4$.    It is easy to get lost in subscripts, so we emphasize that the important feature is that soundness runs in the direction from bits/qubits to checks. This is illustrated in Fig.~\ref{Fig_SingleHP} where it clearly runs the correct way for the classical code but not the quantum code.  For instance, the 2D toric code and 2D Ising code can both be obtained by applying the homological product to a classical repetition code, but only the 2D Ising code exhibits good soundness (recall Fig.~\ref{Fig_ToricIsing}).

Next, we comment on the redundancy of the new quantum code.
\begin{claim}[Updated redundancy]
	\label{redundancyUpdate1}
Let $C_0 \rightarrow_{\delta_0} C_1$ be a chain complex associated with an $[[n,k,d]]$ classical code with check redundancy $\upsilon=n_1/(n_0-k_0)$.  Applying the above homological product we obtain a new chain complex and associated quantum code with check redundancy
\begin{equation}
	\tilde{\upsilon} = \upsilon \frac{ n}{\upsilon(n-k) + k} < 2 \upsilon.
\end{equation}
Notice that if $\upsilon=1$ then $\tilde{\upsilon}=1$.
\end{claim}
To prove this, we begin with the definition of redundancy and apply Eqs.~\eqref{Ntidle} and Eqs.~\eqref{Ktidle}
\begin{align}
\tilde{\upsilon} & =\frac{\tilde{n}_1+\tilde{n}_{-1}}{\tilde{n}_0-\tilde{k}_{0}} \\
& =\frac{2 n_0  n_1 }{n_0^2 + n_1^2 - k_0^2 - k_1^2} \\
& =\frac{2 n_0  n_1 }{(n_0-k_0)(n_0+k_0) + (n_1-k_1)(n_1+k_1) }.
\end{align}
Using that for a length-1  chain complex $n_1-k_1=n_0-k_0$ and the definition of $\upsilon$, we find
\begin{align}
\tilde{\upsilon} &   = \frac{2 n_0  n_1 }{(n_0-k_0)(n_0+k_0+ n_1+k_1) } \\ \nonumber
& = 2 \upsilon \frac{n_0  }{n_0+k_0+ n_1+k_1}   .
\end{align}
Since the fraction is clearly less than 1, we have that $\tilde{\upsilon} < 2 \upsilon$.  Furthermore, using $n_1-k_1=n_0-k_0$ to eliminate $k_1$ and $\upsilon=n_1/(n_0-k_0)$ to eliminate $n_1$, we obtain
\begin{align}
\tilde{\upsilon} &   =  \upsilon \frac{n_0  }{\upsilon (n_0-k_0) + k_0}   ,
\end{align}
and the identification $n=n_0$ and $k=k_0$ gives the final expression for $\tilde{\upsilon}$. 

We conclude this section by considering a simple application of the above homological product.  Given a classical  $[n,k,d]$ code, we can associate many different length-1  chain complexes, depending on whether there is redundancy in the check operators.  However, for any code there always exists a minimal chain complex where there is no redundancy ($\upsilon=1$).  For such a minimal chain complex, we have $n_1 = n - k$, $k_1=0$ and $d_0^T=\infty$.  This is useful as it allows us to make statements that depend only on well known code properties.
\begin{corollary}[Quantum code constructions]
	\label{Cor_QC_constructions}
	Consider a classical  $[n,k,d]$ code.    Applying the above homological product to the minimal chain complex of this code, we obtain a  $[[ 2n(n-k)+k^2 , k^2 , d   ]]$ quantum code with no check redundancy. 
\end{corollary}

\subsection{A second application of the homological product}
\label{Sec_secondHomo}

\begin{figure*}
	\includegraphics{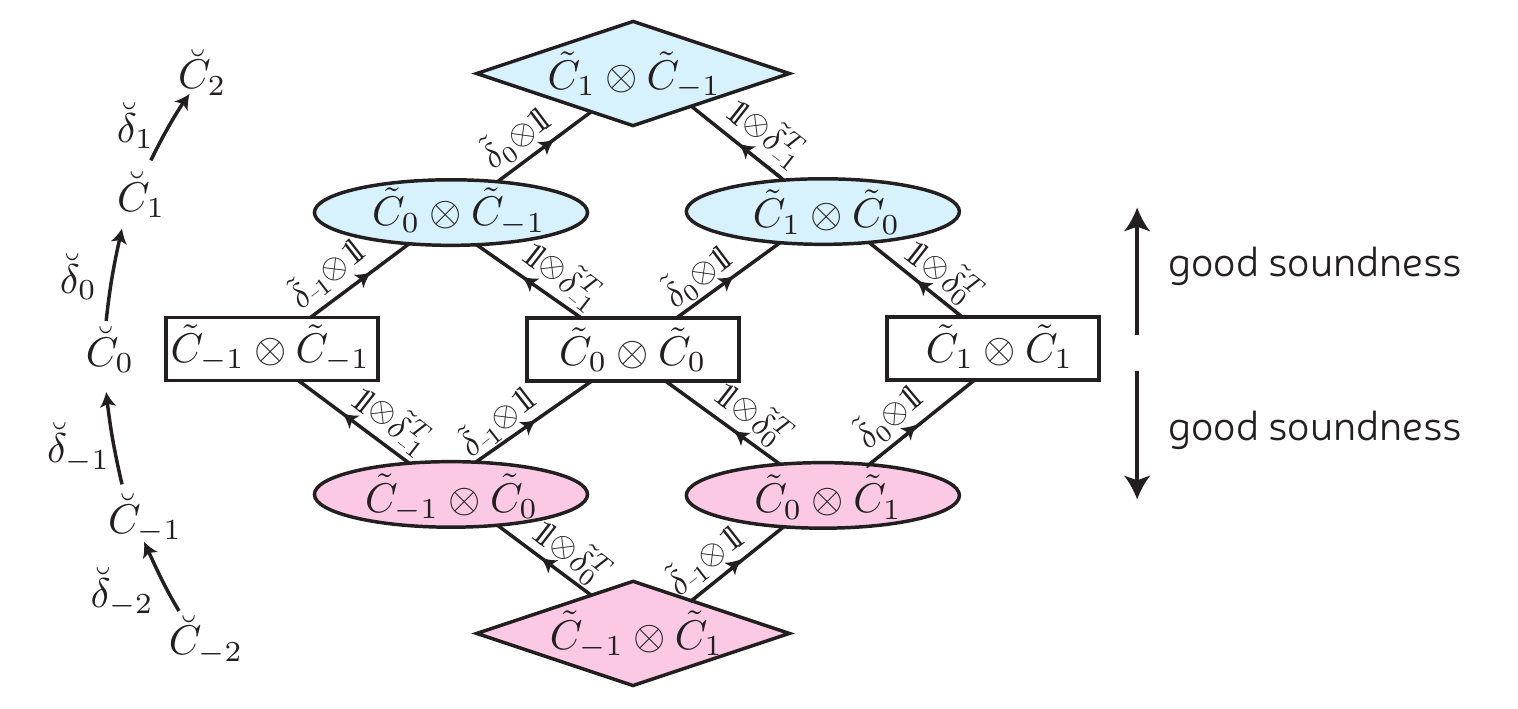}
	\caption{An overview of the second application of the homological product to generate a length-4 chain complex from a two dimensional chain complex (that can be viewed as a quantum code).}
	\label{Fig_DoubleHP}
\end{figure*}

For a quantum error correcting code with metachecks we need a length-4 chain complex, which can be constructed by applying the homological product to a length-2 chain complex.   We use breve ornaments over symbols in this section to identify matrices, variables and vector spaces associated with the length-4 chain complex, as follows
\begin{equation}
\label{fourDimChain}
   \breve{C}_{-2} \rightarrow_{\breve{\delta}_{-2}}     \breve{C}_{-1} \rightarrow_{\breve{\delta}_{-1}}     \breve{C}_{0} \rightarrow_{\breve{\delta}_0}     \breve{C}_{1} \rightarrow_{\breve{\delta}_1}    \breve{C}_{2}  .
\end{equation}
The homological product between a pair of 2-dimensional chain complexes will generate a length-4 chain complex according to the general rule that
\begin{equation}
	 \breve{C}_{m} = \bigoplus_{i-j =m}    \tilde{C_{i}} \otimes \tilde{C_{j}} .
\end{equation}
The boundary maps are illustrated in Fig.~\ref{Fig_DoubleHP} and can be written as block matrices as follows
\begin{align} 
	\breve{\delta}_{-2} & = \left( \begin{array}{c} 
		\id \otimes \tilde{\delta}_0^T  \\
		 \tilde{\delta}_{-1} \otimes \id 
	\end{array} \right) , \\
		\breve{\delta}_{-1} & = \left( \begin{array}{ccc}
	\id \otimes \tilde{\delta}_{-1}^T  & & 0 \\
	\tilde{\delta}_{-1} \otimes \id & & \id \otimes \tilde{\delta}_0^T \\
	 0 & &	 \tilde{\delta}_{0} \otimes \id 
	\end{array} \right)  , \\
		\breve{\delta}_{0} & = \left( \begin{array}{ccccc}
	 \tilde{\delta}_{-1}\otimes \id  & & \id \otimes \tilde{\delta}_{-1}^T & & 0 \\
	0 & &  \tilde{\delta}_0 \otimes \id && \id \otimes \tilde{\delta}_0^T 
	\end{array} \right) , \\
			\breve{\delta}_{1} & = \left( \begin{array}{ccc}
	\tilde{\delta}_{0}\otimes \id  & & \id \otimes \tilde{\delta}_{-1}^T  \\
	\end{array} \right)  .
\end{align}
One can verify that $\breve{\delta}_{j+1}\breve{\delta}_{j}=0$ for all $j$ follows from the same condition on the $\tilde{\delta}$ matrices.  As before, one obtains the relations 
\begin{align}
\label{NandKbreve}
 \breve{n}_m & = \sum_{i-j =m} \tilde{n}_i \tilde{n}_j , \\ \nonumber
  \breve{k}_m & = \sum_{i-j =m} \tilde{k}_i \tilde{k}_j ,
\end{align}
where the first is simple dimension counting and the second line follows from the K\"{u}nneth formula.

The distances are lower bounded as follows
\begin{align}
\label{breveDistances}
\breve{d}_0 , \breve{d}_{-1}^T & \geq \mathrm{min}[ \tilde{d}_{-1} , \mathrm{max}[ \tilde{d}_{0} , \tilde{d}_{-1}^T ] , \tilde{d}_{0}^T  ]  , \\ \nonumber
\breve{d}_{1} , \breve{d}_{-2}^T & \geq \mathrm{min}[ \tilde{d}_{0} , \tilde{d}_{-1}^T   ]  ,
\end{align}
which we prove in App.~\ref{App_distances2}.  Note that the distance will often be significantly larger than these lower bounds.
Our main technical goal is to prove the following soundness result.
\begin{lem}[Second soundness lemma]
	\label{Thm_Coexpand}
	Let $\tilde{C}_{-1} \rightarrow_{\tilde{\delta}_{-1}} \tilde{C}_0 \rightarrow_{\tilde{\delta}_0} \tilde{C}_1$ be a chain complex such that $\tilde{\delta}_{0}^T$ is $(t ,f)$-sound  and $\tilde{\delta}_{-1}$ is $(t ,f)$-sound with $f(x)=x^2/4$.  Applying the above homological product we obtain a new length-4 chain complex (as in Eq.~\ref{fourDimChain}) where the map $\breve{\delta}_{0}$ is $(t ,g)$-sound  and $\breve{\delta}_{-1}^T$ is $(t ,g)$-sound with soundness function $g(x)=x^3/4$.
\end{lem}
We show the direction of the resulting soundness in Fig.~\ref{Fig_DoubleHP} and this should be contrasted with the direction of the soundness arrows in Fig.~\ref{Fig_SingleHP}. We will only prove the results for $\breve{\delta}_{0}$ with the proof for $\breve{\delta}_{-1}^T$ being essentially identical.

Let us first discuss how the problem can be divided into three subproblems.  Let $s \in \im( \breve{\delta}_0)$ so there must exist at least one $r \in \breve{C}_0$ such that $\breve{\delta}_0r=s  $.  We divide $r$ into components 
\begin{equation}
r= \left( \begin{array}{c} r_a \\
r_b \\ r_c 
\end{array} \right) ,
\end{equation}                                        
and consider two distinct images
\begin{align}
\label{Eq_images}
s_{L}  & =  (\tilde{\delta}_{-1} \otimes \id) r_a  + (\id \otimes \tilde{\delta}_{-1}^T)r_b , \\  \label{Eq_images2}
s_{R}  & =  ( \id \otimes \tilde{\delta}_0^T ) r_c + (\tilde{\delta}_{0} \otimes \id) r_b  , 
\end{align}
where
\begin{align}
\label{Eq_imageRelations}
 s = \breve{\delta}_0 r = \left(  \begin{array}{c}
s_{L} \\
s_{R}
\end{array}  \right) .
\end{align}
 One always has the weight relations $|r|=|r_a|+|r_b|+|r_c|$ and $|s|=|s_{L}|+ |s_{R}|$.   
 
For a syndrome that passes all metachecks we have that 
\begin{align}
   \breve{\delta}_1 s & = (\tilde{\delta}_{0} \otimes \id) s_L  +  (\id \otimes \tilde{\delta}_{-1}^T ) s_R  = 0 ,
\end{align}
which entails that
\begin{align}
  m:=(\tilde{\delta}_{0} \otimes \id) s_L & = (\id \otimes \tilde{\delta}_{-1}^T ) s_R  ,
\end{align}
where we have defined this new quantity to be $m$.  Given a physical error pattern $r$ that generates the syndrome (as in Eqs.~\eqref{Eq_images}-\eqref{Eq_images2}) the metachecks are always passed and one finds that  
\begin{align}
\label{mEq}
m = (\tilde{\delta}_0 \otimes \tilde{\delta}_{-1}^T)r_b   .
\end{align}
It is interesting that this depends only on the $r_b$ component of $r$.  We can first try to find low weight $r_b$ that solves Eq.~\eqref{mEq}.   This leads to the following partial solution to the problem
\begin{lem}[Partial soundness result]
	\label{Lem_MiniCoexpand}
	Let $\tilde{C}_{-1} \rightarrow_{\tilde{\delta}_{-1}} \tilde{C}_0 \rightarrow_{\tilde{\delta}_0} \tilde{C}_1$ be a chain complex.  Applying the above homological product we obtain a new length-4 chain complex (as in Eq.~\ref{fourDimChain}) with the following property.   For any $s \in \im ( \breve{\delta}_0)$ there exists an $r_b$ with the following properties
	\begin{enumerate}
		\item correctness: 
		$(\tilde{\delta}_0 \otimes \tilde{\delta}_{-1}^T)r_b = m = (\tilde{\delta}_{0} \otimes \id) s_L  = (\id \otimes \tilde{\delta}_{-1}^T ) s_R$;
		\item low weight: $|r_b| \leq |s_L| \cdot |s_R|$;
		\item  small $s_L$ remainder: $ s_{L} -  (\id \otimes \tilde{\delta}_{-1}^T)r_b = \sum_i \alpha_i \otimes \hat{a}_i$ where $\hat{a}_i$ are unit vectors and $\alpha_i \in \ker{\delta_0}$.  There are at most $|s_L|$ nonzero  $\alpha_i $ and these are bounded in size  $|\alpha_i| \leq |s_L|$ ;
		\item  small $s_R$ remainder: $ s_{R} -  (\tilde{\delta}_{0} \otimes \id) r_b   = \sum_i \hat{b}_i \otimes \beta_i $ where $\hat{b}_i$ are unit vectors and $\beta_i \in \ker{\delta}_{-1}^T$.  There are at most $|s_R|$ nonzero  $\beta_i $ and these are bounded in size  $|\beta_i| \leq |s_R|$.
	\end{enumerate}	
\end{lem}
The proof has a similar flavour to the earlier soundness result and is deferred until App.~\ref{App_Partial_coexpand}.  Notice that the lemma does not require any soundness of the initial chain complex.   Next, we  want to find low-weight $r_a$ and $r_c$ such that they provide the remaining elements of the syndrome as follows
\begin{align}
\label{Eq_remainder}
		(\tilde{\delta}_{-1} \otimes \id) r_a & = s_{L} -  (\id \otimes \tilde{\delta}_{-1}^T)r_b  , \\ 
		( \id \otimes \tilde{\delta}_0^T ) r_c & = s_{R} -  (\tilde{\delta}_{0} \otimes \id) r_b    . 
\end{align}
Fortunately, Lem.~\ref{Lem_MiniCoexpand} ensures that these remainder syndromes are ``small" in the defined sense.  We may next use the  following observation
\begin{claim}[Inheritance of soundness]
	\label{Claim_inherit}
If $\tilde{\delta}_{-1}$ is $(t,f)$-sound then $\tilde{\delta}_{-1} \otimes \id$ is also sound in the following strong sense.  Let $q \in \im ( \tilde{\delta}_{-1} \otimes \id )$ with decomposition $q = \sum_i \alpha_i \otimes \hat{a}_i$ such that $|\alpha_i| < t$ then there exists an $r_a$ such that $(\tilde{\delta}_{-1} \otimes \id)r_a = q$ and $| r_a | \leq \sum_i f( |\alpha_i| )$. A similar result holds when we interchange the order of tensor products and consider $\tilde{\delta}_0^T$.
\end{claim}
The proof is fairly straightforward.   Since $|\alpha_i| < t$ for all $i$ and by assumption $\tilde{\delta}_{-1}$ is $(t,f)$-sound, there must exist $\gamma_i$ such that $\tilde{\delta}_{-1} \gamma_i = \alpha_i$ and $|\gamma_i| \leq f(  |\alpha_i| )$.  By linearity, there exists $r_a=\sum_i \gamma_i  \otimes \hat{a}_i$ such that $(\tilde{\delta}_{-1} \otimes \id)r_a = q$ and $|r_a| \leq \sum_i |\gamma_i | \leq \sum_i f(  |\alpha_i| )$.

Next, we put these pieces together.  Combining Lem.~\ref{Lem_MiniCoexpand} and Claim.~\ref{Claim_inherit} together with the assumption that $|s| < t$ one immediately obtains that there exist $r_a$ and $r_c$ solving Eq.~\eqref{Eq_remainder} with weights upper bounded by 
\begin{align}
    |r_a| & \leq |s_L|   f( |s_L|) \\ 
    |r_c| & \leq |s_R| f( |s_R|)
\end{align}
Therefore, we have the total weight
\begin{align}
|r| & \leq |s_L|   f( |s_L|) +  |s_L| \cdot |s_R|+   |s_R| f( |s_R|) .
\end{align}
We take $f(x)=x^2/4$ as stated in Thm.~\ref{Thm_Coexpand}, which leads to
\begin{align}
|r|& \leq \frac{1}{4}|s_L|^3 +  |s_L| \cdot |s_R|+  \frac{1}{4} |s_R|^3 \\
&\leq \frac{1}{4}( |s_L| + |s_R|)^3  \\
&= \frac{1}{4} |s|^3  .
\end{align}
Therefore, we have proven ($t,g$)-sound of $\breve{\delta_0}$ with $g(x)=x^3/4$.  This completes the proof that Thm.~\ref{Thm_Coexpand} follows from Lem.~\ref{Lem_MiniCoexpand}. 

Next, we comment on the check redundancy of these codes
\begin{claim}[Updated redundancy part 2]
	\label{redundancyUpdate2}
	Consider a length-2 chain complex and associated quantum code with check redundancy $\tilde{\upsilon}$.  Applying the above homological product we obtain a length-4 chain complex and new quantum code with check redundancy $\breve{\upsilon} < 2 \tilde{\upsilon}$.
\end{claim}
To prove this we recall the definition of redundancy and then use Eqs.~\eqref{NandKbreve} to obtain
\begin{align}
	\breve{\upsilon} & = \frac{\breve{n}_1 + \breve{n}_{-1}}{\breve{n}_0 - \breve{k}_0 } \\
	&= \frac{2 \tilde{n}_0  (\tilde{n}_1 + \tilde{n}_{-1}) }{ (\tilde{n}^2_{-1} + \tilde{n}^2_0+  \tilde{n}^2_{1}  ) - (\tilde{k}^2_{-1} + \tilde{k}^2_0+  \tilde{k}^2_{1}  )   }  .
\end{align}
Since $\tilde{n}_j \geq \tilde{k}_j$ for all $j$, the denominator is greater than $\tilde{n}^2_0-\tilde{k}^2_0$, which itself can be factorised as $(\tilde{n}_0-\tilde{k}_0)(\tilde{n}_0+\tilde{k}_0)$ and so
\begin{align}
\breve{\upsilon} & \leq \frac{2 \tilde{n}_0 (\tilde{n}_1 + \tilde{n}_{-1}) }{ (\tilde{n}_0-\tilde{k}_0)(\tilde{n}_0+\tilde{k}_0)  }  , \\
& = 2 \left(\frac{\tilde{n}_1 + \tilde{n}_{-1} }{ \tilde{n}_0-\tilde{k}_0 } \right)  \left( \frac{\tilde{n}_0}{ \tilde{n}_0+\tilde{k}_0  } \right) , \\
& = 2 \tilde{\upsilon} \left(\frac{ \tilde{n}_0 }{ \tilde{n}_0+\tilde{k}_0 }   \right) ,
\end{align} 
Last, we use the loose bound that the fraction is less than 1  to conclude that $\breve{\upsilon} \leq 2 \tilde{\upsilon} $ as claimed.

\subsection{Combining homological products}

Here we combine the results of the preceding two subsections.  Parameters carrying a breve are first expressed in term of parameters carrying a tilde, and then the tilde parameters are replaced with unornamented parameters.  
\begin{align}
	\breve{n}_{0} &  = \tilde{n}^2_1+ \tilde{n}_0^2 +  \tilde{n}^2_{-1} = (n_0^2+ n_1^2)^2 + 2n_0^2n_1^2 ,  \\ \nonumber
	\breve{n}_{1} = \breve{n}_{-1} &  	 = \tilde{n}_0(\tilde{n}_1+ \tilde{n}_{-1}) = 2 (n_0^2+n_1^2)n_0 n_1 , \\ \nonumber
	\breve{k}_{0} &  = \tilde{k}^2_1+ \tilde{k}_0^2 +  \tilde{k}^2_{-1} = (k_0^2+ k_1^2)^2 + 2k_0^2k_1^2 ,  \\ \nonumber
	\breve{k}_{1} = 		\breve{k}_{-1}  &   = \tilde{k}_0(\tilde{k}_1+\tilde{k}_{-1})= 2 (k_0^2+k_1^2)k_0 k_1 ,  \\ \nonumber
	\breve{d}_{0} = \breve{d}_{-1}^T &   \geq \mathrm{min}[  d_0 , d_0^T ] ,  \\ \nonumber
	\breve{d}_{1} = 	\breve{d}_{-1}^T &  \geq \mathrm{min}[  d_0 , d_0^T ]  .  
\end{align}	
Furthermore, by combining Claim.~\ref{redundancyUpdate1} and Claim.~\ref{redundancyUpdate2} we obtain an upper bound on the check redundancy
\begin{align}
\breve{\upsilon} < 2 \tilde{\upsilon}= 2 \upsilon \frac{ n}{\upsilon(n-k) + k}  ,
\end{align}	
where $ \upsilon$ is the check redundancy of the $[n,k,d]$ classical code associated with the initial length-1  chain complex.

The simplest case is when we use a minimal chain complex representing the initial $[n,k,d]$ classical code.  Then $\upsilon=1$, $k_1=0$ and $n_1=n-k$ and the above equations simplify to 
\begin{align}
\breve{n}_{0} &  = n^4 + 4 n^2 (n-k)^2 + (n-k)^4 ,  \\ \nonumber
\breve{n}_{1} = \breve{n}_{-1} &  	 = 2n (n-k) (n^2+(n-k)^2) , \\ \nonumber
\breve{k}_{0} &  =k^4,  \\ \nonumber
\breve{k}_{1} = 		\breve{k}_{-1}  &   = 0  \\ \nonumber
\breve{\upsilon} & < 2. \\ \nonumber
	\breve{d}_{0} = \breve{d}_{-1}^T &   \geq d ,  
\end{align}	
We  also know that $\breve{d}_{1} = 	\breve{d}_{-2}^T  = \infty$ as a consequence of $\breve{k}_{1} = 		\breve{k}_{-1}     = 0$.  We make the following identifications:  $\breve{n}_{0}$ gives the number of physical qubits $n_Q$; $\breve{k}_{0}$ is the number of logical qubits $k_Q$; $	\breve{d}_{0}   $ and 	$\breve{d}_{-1}^T$ give the qubit error distance $d_Q$; and $\breve{d}_{1}$ and $\breve{d}_{-2}^T$ give the single-shot distance $d_{ss}$.  This proves Thm.~\ref{THM_construct}.  We remark that in the final stages of this research, Zeng and Pryadko posted a preprint~\cite{zeng2018higher} that shows that the distance is much better than suggested by our bounds, in particular $\breve{d}_{0} = \breve{d}_{-1}^T  = d^2 $.  
 
 	\begin{table*}
 	\centering
 	\begin{tabular}{|c|c|c|c|c|c||c|c|c|c|c|c|c|}
 		\multicolumn{6}{|c||}{\textbf{Input classical code}}  & 	 \multicolumn{7}{|c|}{\textbf{Double homological product code}} \\ &
 		\multicolumn{3}{|c|}{parameters} & max. check & redundancy & \multicolumn{4}{c|}{parameters}  & max. check & mean check & redundancy \\
 		$\delta$ & $n$ & $k$ & $d$ &   weight & $\upsilon$ & $n_Q$ & $k_Q$ & $d_Q$ & $d_{ss}$ &  weight &  weight & $\breve{\upsilon} $ \\ \hline
 		$\left( \begin{array}{ccc}
 		1 & 1 & 0 \\
 		0 & 1 & 1 \\
 		\end{array} \right)$ & 3 & 1 & 3 & 2 & 1 & 241 & 1 &  9 & $\infty$ & 6 & 4.87179 & 1.3  \\
 		$\left( \begin{array}{cccc}
 		1 & 1 & 0 & 0 \\
 		0 & 1 & 1 & 0 \\
 		0 & 0 & 1 & 1 \\
 		\end{array} \right)$ & 4 & 1 & 4 & 2 & 1 & 913 & 1 &  16 & $\infty$  & 6 & 5.18 & 1.31579  \\
 		$\left( \begin{array}{ccc}
 		1 & 1 & 0 \\
 		0 & 1 & 1 \\
 		1 & 0 & 1 \\
 		\end{array} \right)$ & 3 & 1 & 3 & 2 & 1.5 & 486 & 6 &  9 & 3 & 6 & 6 & 1.33884  \\		
 		$\left( \begin{array}{cccccc}
 		1 & 1 & 0 & 0 & 0 & 0 \\
 		0 & 1 & 1 & 0 & 1 & 0 \\
 		0 & 0 & 1 & 1 & 0  & 0 \\
 		0  & 0 & 0  & 0 & 1 & 1 \\
 		\end{array} \right)$ & 6 & 2 & 4 & 3 & 1 & 3856 & 16 & 16 & $\infty$ & 8 & 5.48077 & 1.3  \\				
 	\end{tabular}
 	\caption{Some example small classical codes used to generate a quantum code with good soundness through a double application of the homological product. Many of the parameters come directly from equations in the main text.  The mean check weight and redundancy are calculated exactly by constructing the explicit parity check matrices.  Our table uses the improved distance $d_Q$ results of Zeng and Pryadko~\cite{zeng2018higher}.} \label{tab_examples}
 \end{table*}

 In Table~\ref{tab_examples} we provide some concrete examples.  These are the smallest examples since they use very small initial classical codes.  Though the resulting quantum code is much larger.  The first three examples correspond to 4D toric codes with cubic tilling either with closed boundary conditions (examples 1 and 2) or periodic boundary conditions (example 3). The last example corresponds to no previous codes that we know of.  We have deliberately chosen codes that have low check weight as these will be the most experimentally feasible.  Our constructions could potentially be slightly improved using a generalisation of the hypergraph improvements analogous to use of rotated toric lattices~\cite{kovalev12}.
 
 \section{Discussion \& Conclusions}
 
This is a paper of two halves.  The first half was conceptual and gave a presentation of single-shot error correction. We found an  intimate connection between single-shot error correction and a property called good soundness.   We saw that good soundness in LDPC codes entails a macroscopic energy barrier, which further confirms a relationship between passive quantum memories and single-shot error correction.  However, our results leave open whether there exist any codes with a macroscopic energy barrier that lack good soundness. Michael Beverland suggested in discussion that it would be interesting to look at whether Haah's cubic code~\cite{haah2011cubic,bravyi2013quantum} has good soundness.  The Haah cubic code is notable because it does have a macroscopic energy barrier but is not a good passive quantum memory at all scales due to entropic effects.  Also curious is the role of metachecks and redundancy.  We saw that good soundness can be achieved by any code without any check redundancy, but the proof used a diagonalised form of the stabiliser generators that typically destroys any LDPC properties.  

The second half of this paper was more technical and focused on specific code constructions capable of providing both good soundness and LDPC properties.  It has long been known that homology theory provides a natural mathematical framework for CCS codes, but we saw that homology theory is especially useful when metachecks (checks on measurements) are added to the picture.  It is well known that for topological codes the energy barrier and single-shot error correction are intimately related to the dimensionality of the code.  We abstract away the topological structure and instead work with algebraic homological structure.  While these codes no longer have a dimensionality in the geometric sense, we saw that using the homological product can imbue codes with a sort of effective dimensionality.  More precisely, a double application of the homological product resulted in single-shot properties similar to 4-dimensional topological codes. Many readers will feel more comfortable with topological codes because of the conceptual and visual crutches they provide.  However, topological codes are significantly limited in terms of the code parameters they can achieve due to trade-off bounds~\cite{bravyi2010tradeoffs,delfosse2013tradeoffs}.  So by freeing ourselves from the constraints of topological codes and pursuing their more abstract cousins, we can seemingly benefit from many of the advantages of high-dimensional topological codes (e.g. single-shot error correction) but with improved code parameters.   This prompts the question what other topological code properties might hold for homological product codes.  We know that 3D and 4D topological codes can support transversal non-Clifford gates~\cite{bombin2006topological,bombin2013self,watson2015qudit,kubica2015unfolding,kubica2015universal,vasmer2018universal,campbell2017roads}, which suggests that a similar property might hold for suitably defined homological product codes.  

Our code constructions married  good soundness and LDPC properties, through the use of check redundancy and associated metachecks.  But do any codes exist without check redundancy that are useful for single-shot error correction?  A related question is whether our soundness properties are necessary conditions for single-shot error correction as well as being sufficient conditions. While finishing this research, work on quantum expander codes~\cite{leverrier18} has shown that they can perform single-shot error correction without any check redundancy.   Initially, we speculated (in an early preprint) that the quantum error codes will have good soundness, but Leverrier has shown (in private correspondence) that they do not have this property!  Therefore, there is more work to be done on this topic to find a code property more permissive than soundness that encompasses all of our codes and also the quantum expander codes.
 
 The main limitation of this work is that we restrict our attention to adversarial noise.   Stochastic noise models instead distribute errors according to some probability distribution and assign a non-zero probability to every error configuration.  If the probability of a high weight error is low, then we can still leverage proofs from the adversarial noise setting.  However, in an independent noise model where each qubit is affected with probability $p$, a code with $n$ qubits will typically suffer around $pn$ errors.  For all known quantum LDPC code families, the distance scales sublinearly, and so there is some scale at which the code is likely to suffer an error considerable larger than the code distance.   Nevertheless, one is often able to prove the existence of an error correcting threshold.  The crucial point is that even though some errors of weight $pn$ might not be correctable, these represent a small fraction of all weight $pn$ errors and so happen with small probability.  At this point, proof techniques diverge. We can prove that this works for concatenated codes, topological codes and low-density parity check codes~\cite{kovalev2013fault}.  As such, while there is a single theoretical framework for adversarial noise, there is no single theory for stochastic noise in all settings.  The situation is likely the same in the setting of single-shot error correction.   The pioneering work of Bombin demonstrated that three dimensional colour codes can perform single-shot error correction against a stochastic noise model~\cite{bombin2015single}, and so in this sense our results are strictly weaker.  On the other hand, our approach is strictly more general as it applies to a broad range of codes, including many new code constructions such as those presented here.  It is then natural to wonder what are sufficient and necessary conditions for single-shot error correction to work against stochastic noise?  It is reasonable to conjecture that any concatenated or LDPC codes that meets our criteria for adversarial noise will also perform single-shot error correction against stochastic noise.  
 
Acknowledgements.-  This work was supported by the EPSRC (EP/M024261/1) and the QCDA project which has received funding from the QuantERA ERA-NET Cofund in Quantum Technologies implemented within the European Union's Horizon 2020 Programme.  I would like to thank Nicolas Delfosse for his tutorial on hypergraph product codes during the FTQT 2016 workshop at the Centro de Ciencias de Benasque Pedro Pascual.  Thank you to Simon Willerton, Michael Beverland, Mike Vasmer, Anthony Leverrier, Barbara Terhal and Ben Brown for conservations and comments on the manuscript.  Referee 2 is thanked for their diligent attention to detail.


\appendix

\section{A simple proof of relation between Betti numbers}
\label{App_Poincare}

We give a simple proof that $k_j = k_j^T$ as defined in Eq.~\eqref{Eq_Betti} and Eq.~\eqref{Eq_CoBetti}.  The proof uses simple linear algebra rather than sophisticated homological techniques that are needed in more exotic settings.   We use the rank-nullity theorem that for any matrix $A$,
\begin{equation}
	\mathrm{rank}(A)+	\mathrm{nullity}(A) = n ,
\end{equation}
where $n$ is the number of columns in $A$.  This entails that
\begin{align}  \label{RankNull}
\mathrm{rank}(\delta_j)+	\mathrm{nullity}(\delta_j) & = n_j , \\  
\mathrm{rank}(\delta_{j-1}^T)+	\mathrm{nullity}(\delta_{j-1}^T) & = n_j . \label{coRankNull}
\end{align}
Taking the definition of $k_j^T$ (recall Eq.~\eqref{Eq_CoBetti}) and using Eq.~\eqref{coRankNull} to eliminate the dependence on $\mathrm{nullity}(\delta_{j-1}^T)$, we obtain
\begin{equation}
   k_j^T = n_j -  \mathrm{rank}(\delta_{j-1}^T) -  \mathrm{rank}(\delta_{j}^T).
\end{equation}
Using that for any matrix  $\mathrm{rank}(A)=\mathrm{rank}(A^T)$, we deduce
\begin{equation}
k_j^T = n_j -  \mathrm{rank}(\delta_{j-1}) -  \mathrm{rank}(\delta_{j}).
\end{equation}
Using Eq.~\eqref{RankNull} to eliminate $\mathrm{rank}(\delta_{j})$, we get
\begin{align}
k_j^T & = n_j -  \mathrm{rank}(\delta_{j-1})   -  [n_j - \mathrm{nullity}(\delta_{j}) ] \\ \nonumber
& =  \mathrm{nullity}(\delta_{j})   - \mathrm{rank}(\delta_{j-1}) , 
\end{align}
which is precisely the definition of $k_j$ given in Eq.~\eqref{Eq_Betti}.  This completes this simple but educational proof.

\section{Further notation}

\subsection{Vector reshaping}
\label{App_Matrix_reshaping}

Throughout the appendices we often reshape vectors into matrices.  If we have a vector $v$ belonging to some tensor product space $A \otimes B$, then we can reshape $v$ into a matrix $V$.  We always use lower-case symbols for vectors and upper-case for the resulting matrix after reshaping.  Let $\{ \hat{a}_i \}$ and $\{ \hat{b}_j \}$ be unit basis vectors for $A$ and $B$, respectively.  Then any vector $v$ can be decomposed in this basis as
\begin{equation}
	 v = \sum_{i,j} V_{i,j} \hat{a}_i \otimes \hat{b}_j ,
\end{equation}	
where the coefficients $V_{i,j}$ are elements of the matrix representation. That is, $V_{i,j}$ is the entry in the $i^{\mathrm{th}}$ row and $j^{\mathrm{th}}$ column of matrix $V$.  Furthermore, given matrices $M: A \rightarrow A$ and $N: B \rightarrow B$ we will rewrite equations as follows 
\begin{equation}
 (M \otimes N)	v \rightarrow M V N^T ,
\end{equation}
which is easily verified.

\subsection{Matrix support}
\label{App_Matrix_support}

We further introduce the notion of column and row support.  Given any matrix $X$ we let $\mathrm{colsupp}(X)$ denote the set of columns in $X$ with at least one nonzero entry.
 Given any matrix $X$ we let $\mathrm{rowsupp}(X)$ denote the set of rows in $X$ with at least one nonzero entry. We shall often use $| \ldots |$ to denote the number of rows or columns within some support.  That is, $|\mathrm{colsupp}(X)|$ is the number of columns in $X$ with at least one nonzero entry. For example, if
\begin{equation}
	X = \left( \begin{array}{cccccc}
		1 & 0 & 0 & 1 & 1  & 0 \\ 
		0 & 1 & 0 & 1 & 1 & 0 \\ 
		0 & 0 & 0 & 1 & 1 & 0 \\ 
	\end{array}		
	\right),
\end{equation}	
then $\mathrm{colsupp}(X)=\{ 1,2,4,5 \}$ and $\mathrm{rowsupp}(X)=\{ 1,2,3 \}$.  Furthermore, $|\mathrm{colsupp}(X)|=4$ and $|\mathrm{rowsupp}(X)|=3$.

\section{Distance bounds: part one}
\label{App_distances}

Here we give proofs of distances associated with length-2 chain complexes constructed using the homological product (see Eqs.~\eqref{Eq_distances1}-\eqref{Eq_distances4}).  

\subsection{First bound}
\label{Sec_FirstBound}

We begin by showing that $\tilde{d}_{-1}  \geq  d_0 d_0^T$.   The quantity $\tilde{d}_{-1}$ is the weight of the smallest nonzero vector $r \in C_0 \otimes C_1$ such that $\tilde{\delta}_{-1}r = 0$.   We use that $r \in C_0 \otimes C_1$ can be reshaped into a matrix $R$;  see \ref{App_Matrix_reshaping} for discussion of reshaping.  The condition $\tilde{\delta}_{-1} r $ entails that every column of $R$ must be in $\mathrm{ker}(\delta_0)$ and every row of $R$ must be in $\mathrm{ker}(\delta_0^T)$.  Assuming, $R$ is nonzero, there must be at least one non-zero column.  Since this column has weight at least $d_0$, it follows that there are at least $d_0$ non-zero rows. Each of these rows has weight at least $d_0^T$.  Therefore, the total weight is at least $d_0 d_0^T$ as required.   Next, we show $\tilde{d}_{-1}  \leq  d_0 d_0^T$.  We assume, $ d_0 \neq \infty $ and $ d^T_0 \neq \infty$ otherwise the inequality is trivially true.  Let $\alpha$ be a minimal weight non-zero vector in the kernel of $\delta_0$, so $|\alpha|=d_0$. Similarly let $\beta \in \ker (\delta_0^T)$ with $|\beta| = d_0^T$.  Then $\alpha \otimes \beta \in C_0 \otimes C_1$ has $|\alpha \otimes \beta| = d_0 d_0^T$ and is easily verified to satisfy $\tilde{\delta}_{-1}(\alpha \otimes \beta)=0$. The proof of $\tilde{d}_0^T  = d_0 d_0^T$ follows by symmetry. 

\subsection{Second bound}
\label{Sec_SecondBound}

Next we show that $\tilde{d}_{0} \geq  \mathrm{min}[d_0 , d_0^T]$.  Recall, this is the  weight of the smallest vector $r$ such that $\tilde{\delta}_{0}r = 0$ and $r \notin \im ( \tilde{\delta}_{-1} )$.  All $r$  can be decomposed as
\begin{equation}
	r = \left( \begin{array}{c}
	r_a \\
	r_b
	\end{array} \right) ,
\end{equation}
where $\tilde{\delta}_{0}r = 0$ entails that $(\delta_0 \otimes \id)r_{a}=(\id \otimes \delta_0^T )r_{b}$.  Assuming $r$ is a non-trivial cycle, it follows that there must exist a cocycle $w=(w_a, w_b)$ such that $w^T r=1$. Therefore,  $w_a^Tr_a + w_b^T r_b =1$ and either $w_a^T r_a=1$ or $w_b^T r_b=1$. We proceed assuming $w_a^Tr_a  =1$ and further note that the cocycle can always be assumed to have the form $w= (e \otimes f) \oplus 0$. This is a good place to remind the reader that $\oplus$ is the direct product and when applied to columns vectors means that we stack the columns.  Since $w$ ought to be a cocycle it must satisfy $\tilde{\delta}_{-1}^T w = 0$ which entails that $\delta_0 f =0$.  The relation $w^T r=1$ then becomes $(e^T \otimes f^T) r_a = 1$.  We can reshape some vectors into matrices, and these equations become
\begin{align}
	(e^T \otimes f^T) r_a = 1	& \implies  e^T R_a f = 1 \\ 
(\delta_0 \otimes \id)r_{a}=(\id \otimes \delta_0^T )r_{b} & \implies \delta_0 R_a = R_b \delta_0 
\end{align}	
We consider the vector $R_a f$.  From $ \delta_0 R_a = R_b \delta_0 $ we infer that $\delta_0 (R_a f) =  R_b \delta_0 f$.  Using also that $\delta_0 f=0$ we have a proof that $\delta_0 (R_a f) =0$ and so $R_a f \in \ker (\delta_0)$.  However, $R_a f \neq 0$ otherwise it would be impossible to satisfy $e^T R_a f = 1 $.  It follows that $d_0 \leq |R_a f|$. Since $R_a f$ is formed from linear combinations of columns from $R_a$, we have $|R_a f| \leq |R_a|$ and hence $d_{0} \leq  |R_a|$.  It follows that $d_{0} \leq  |r| $ in this case.   For the $w_b^T r_b=1$ case, a similar argument follows but giving a lower bound of $d_0^T \leq |r|$.  Therefore, the actual lower bound on $|r|$ is the minimum of these two cases. 

\section{Soundness proof: part one}
\label{App_Coexpand1}

Here we prove Lem.~\ref{Lem_Coexpand} for $\tilde{\delta}_{0}^T$, with the $\tilde{\delta}_{-1}$ proof following a similar fashion.  Recalling the definition of soundness, we consider $s \in \tilde{C}_0$ such that $s \in \im(\tilde{\delta}_0^T)$ and $|s| < t = \mathrm{min}(d_0^T, d_0)$.  Therefore, both $|s|< d_0^T$ and $|s|< d_0$ hold.  There must exist at least one $r \in  \tilde{C}_{1} = C_{1} \otimes C_{0} $ such that $s=\tilde{\delta}_{0}^T r$.  This will not be the only possible solution, but let us begin by exploring the relationship between $|s|$ and $|r|$.  

The vector $s$ has two components $s=s_L \oplus s_R$ and breaking $s=\tilde{\delta}_{0}^T r$ into components, we have 
\begin{align}
  s_L &  =(\delta_0^T \otimes \id) r ,  \\ \nonumber
  s_R &  =(\id \otimes \delta_0) r .
\end{align}
Next, we reshape $r$, $s_L$ and $s_R$ into matrices (see \ref{App_Matrix_reshaping} for discussion of reshaping) so that 
\begin{align}
	\label{SR_relations}
s_L   =(\delta_0^T \otimes \id) r  & \implies S_L =	\delta_0^T R  , \\ \nonumber
s_R   =(\id \otimes \delta_0) r   & \implies S_R = R 	\delta_0^T .
\end{align}
In terms of support (recall notation of App.~\ref{App_Matrix_support}) the above equations entail that 
\begin{align}
	\colsupp (  S_L )  & \subseteq  	\colsupp (  R ) , \\ \nonumber
	\rowsupp (  S_R )  & \subseteq  	\rowsupp ( R )	.
\end{align}
In general, this means that
\begin{align}
	\label{Maybe_equality1}
	| \colsupp (  S_L ) |  & \leq  | \colsupp (  R ) |  ,\\ 	\label{Maybe_equality2}
	| \rowsupp (  S_R ) | & \leq  |	\rowsupp ( R   )	| .
\end{align}
Using $|X|$ to denote the number of 1s contained in a binary matrix $X$, we remark that $| \colsupp (  X ) | \leq |X|$ and $| \rowsupp (  X ) | \leq |X|$ for any $X$, and so
\begin{align}
	| S_L  |  & \geq  | \colsupp (  S_L ) |  ,\\ \nonumber
	|   S_R  | & \geq  | \rowsupp ( S_R   )	 | .
\end{align}
Combined with $|s|=|   S_L  |+|S_R| $ we find
\begin{align}
	\label{s_col_row_supp}
	|s|  & \geq  | \colsupp (  S_L ) |  +  | \rowsupp ( S_R   )	 | .
\end{align}
Squaring both sides and using $(a+b)^2/ 4 \geq  ab$ for integer $a$ and $b$, we obtain 
\begin{align}
	\label{Eq_s_bound}
	|s|^2 / 4 & \geq  | \colsupp (  S_L ) | \cdot  | \rowsupp ( S_R   )	 | .
\end{align}
We would like to substitute in Eqs.~\eqref{Maybe_equality1}-\eqref{Maybe_equality2} but the inequality signs do not align correctly.  We would be able to proceed if Eqs.~\eqref{Maybe_equality1}-\eqref{Maybe_equality2} held with strict equality, but this is not always the case. 

\begin{figure*}[t]
	\includegraphics{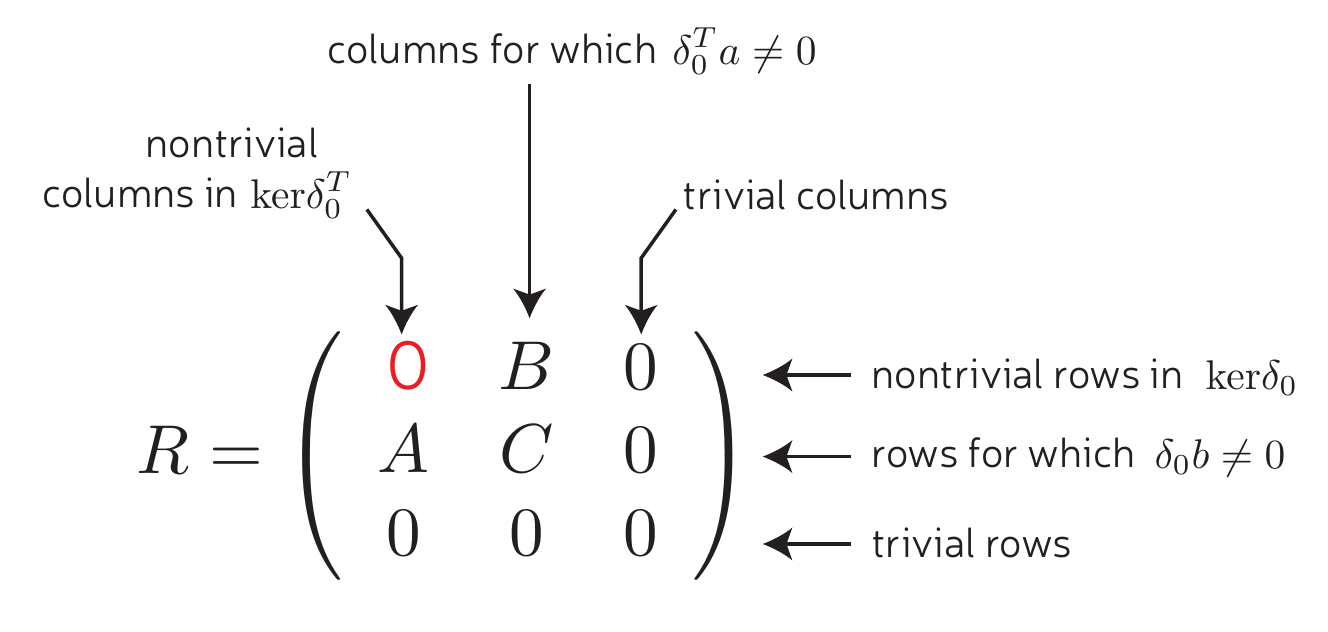}
	\caption{The form of $R$ after the repeated $R \rightarrow R + a b^T$ process has terminated. We have taken the liberty of permuting columns and rows, such that: any column of $R$ in $\ker(\delta_0)$ will intersect block $A$; any row (transposed) of $R$ in $\ker(\delta^T_0)$ will intersect block $B$.  Since the aforementioned $R \rightarrow R + a b^T$ process has terminated, there are no more column and row pairs such that they are in the relevant kernel and they intersect.  Therefore, the upper-left block must be all-zero as shown otherwise there would still exist such an intersecting pair and the $R \rightarrow R + a b^T$ process ought to continue.  Note further that the process must terminate after a finite number of rounds since the column and row supports are strictly decreasing with each transform of $R$.		Since the middle block of columns are those that do not vanish under $\delta_0^T$, we have that $|S_L| \geq |\delta_0^TR|$ is equal to the number of columns in the middle block of columns.  Similarly, $|S_R|$ is lower bounded by the number of rows in the middle block of rows.}
	\label{Fig_Rmat}
\end{figure*}

To proceed we use that the above $R$ is not the only possible solution.   Given an initial $R$ we can transform to obtain a new $R$ so that Eqs.~\eqref{SR_relations} are preserved, but so that also Eq.~\eqref{Maybe_equality1} and Eq.~\eqref{Maybe_equality2} become equalities. In particular, given a pair of vectors $a \in \ker{\delta_0^T}$ and $b \in \ker{\delta_0}$ we can perform $R \rightarrow R + a b^T$ and Eqs.~\eqref{SR_relations} will be preserved.  We assume for now that neither Eq.~\eqref{Maybe_equality1} nor Eq.~\eqref{Maybe_equality2} are strict equalities, and so we may take both $a$ and $b^T$ to be column and row vectors from $R$.   It follows that the new $R + a b^T$ has column and row support strictly contained within that of $R$, and the support may even reduce in size.  Notice that adding $a b^T$ will add $a$ to every column in $R$ on which $b^T$ is supported.  So if the support of $a$ and $b^T$ intersect in $R$, we can strictly decrease the number of columns and the number of rows in $R$.  By intersect in $R$ we mean that if $b$ is the $i^{\mathrm{th}}$ row of $R$ and $a$ is the $j^{\mathrm{th}}$ column of $R$, then $R_{i,j}=1$.  Let us consider an example,
\begin{equation}
	R = \left( \begin{array}{ccccc}
		0 & 0 & 0 & 1 & 0 \\ 
		0 & 0 & 0 & 1 & 0 \\ 
        0 & 0 & 0 & 1 & 0 \\ 
        0 & 0 & 0 & 1 & 0 \\ 
        1 & 1 & 1  & \textbf{\textcolor{red}{1}} & 0 \\ 		
		\end{array}		
		  \right),
\end{equation}	
so that $\mathrm{colsupp}(R)=\{ 1,2,3,4  \}$ and $\mathrm{rowsupp}(R)=\{ 1,2,3,4,5  \}$.  Let $a=(1,1,1,1,1)^T$ be the fourth column vector and $b=(1,1,1,1,0)$ be the last row vector.  They intersect since $R_{5,4}=1$ and we emphasis this by highlighting the intersecting element in bold and red.  We find that 
\begin{equation}
	R' =R + a b^T  = \left( \begin{array}{ccccc}
		1 & 1 & 1 & 0 & 0 \\ 
		1 & 1 & 1 & 0 & 0 \\ 
		1 & 1 & 1 & 0 & 0 \\ 
		1 & 1 & 1 & 0 & 0 \\ 
		0 & 0 & 0 & 0 & 0 \\ 		
	\end{array}		
	\right).
\end{equation}	
Notice that $\mathrm{colsupp}(R)=\{ 1,2,3  \}$ and $\mathrm{rowsupp}(R)=\{ 1,2,3,4  \}$, so that the supports have strictly decreased.  Also note that the intersection property was crucial.  If we had instead considered
\begin{equation}
	R = \left( \begin{array}{ccccc}
		0 & 0 & 0 & 1 & 0 \\ 
		0 & 0 & 0 & 1 & 0 \\ 
		0 & 0 & 0 & 1 & 0 \\ 
		0 & 0 & 0 & 1 & 0 \\ 
		1 & 1 & 1  & \textbf{\textcolor{red}{0}} & 0 \\ 		
	\end{array}		
	\right),
\end{equation}	
with non-intersecting  $a=(1,1,1,1,0)^T$ and $b=(1,1,1,0,0)$ then we would find
\begin{equation}
	R' =R + a b^T  = \left( \begin{array}{ccccc}
		1 & 1 & 1 & 1 & 0 \\ 
		1 & 1 & 1 & 1 & 0 \\ 
		1 & 1 & 1 & 1 & 0 \\ 
		1 & 1 & 1 & 1 & 0 \\ 
		1 & 1 & 1 & 0 & 0 \\ 		
	\end{array}		
	\right).
\end{equation}	
The column and row support is completely unchanged.  The key point is that when $a$ and $b^T$ intersect in $R$, we will add $a$ to a set of columns including the column equal to $a$.  Since we do this modulo 2, at least one column is removed.  Similarly, at least one row will be removed.

Repeating this $R \rightarrow R + a b^T$ process must terminate when there are no remaining column/row pairs that intersect and are elements of the relevant kernels.   After termination the matrix $R$ was a special form best illustrated using a block matrix equation shown in Fig.~\ref{Fig_Rmat} with further comment in the figure caption.   Having transformed into this special form, we next use additional assumptions under which $A$ and $B$ blocks vanish and so Eq.~\eqref{Maybe_equality1}  and Eq.~\eqref{Maybe_equality2} become strict equalities.  Assume $A$ is nonzero so there exists a column vector $c$ intersecting block $A$.  Since $c \in \ker(\delta^T_0)$ and $c \neq 0$ we have $|c| \geq  d_0^T $.  Furthermore, since column vector $c$ intersects block $A$ we conclude that the middle block of rows in $R$ must contains at least $|c|$ nonzero rows  and consequently, $| \rowsupp (  S_R ) |  \geq |c|$ (see Fig.~\ref{Fig_Rmat} and caption for more intuition) and consequently $| \rowsupp (  S_R ) |  \geq d_0^T $. Next, we use our assumption that $|s| < d_0^T$ that was asserted at the very start of this proof, which we combine with Eq.~\eqref{s_col_row_supp} to conclude that
\begin{equation}
	 d_0^T >  | \colsupp (  S_L ) |  +  | \rowsupp ( S_R   ) | ,
\end{equation}	
and so  $d_0^T >  | \rowsupp (  S_R ) | $.      Having proved both $| \rowsupp (  S_R ) |  \geq d_0^T $ and  $d_0^T >  | \rowsupp (  S_R ) | $.  We have a contradiction that is only resolved if such a column vector $c$ does not actually exist and therefore $A=0$.

 Using a nonzero row vector in $\mathrm{ker}(\delta_0)$, a similar argument entails that $| \colsupp (  S_L ) |  \geq d_0 $, which contradicts $|s|< d_0$, and so we conclude $B=0$ also. Therefore, we see that the above transformation must yield a form with where Eqs.~\eqref{Maybe_equality1}-\eqref{Maybe_equality2} hold with strict equality.  This can be combined with Eq.~\eqref{Eq_s_bound} to conclude that
\begin{align}
	\label{Eq_s_bound2}
	|s|^2 / 4 & \geq  | \colsupp (  R ) | \cdot  | \rowsupp ( R  )	 | .
\end{align}
Furthermore, if $R$ is supported on a submatrix of size $ | \colsupp (  R ) | $ by $ | \rowsupp ( R   )	 | $ then the size of this submatrix gives an upper bound on $|R|=|r|$ so that 
\begin{align} \label{Eq_r_bound}
 | \colsupp (  R ) |  \cdot  | \rowsupp ( R   )	 |  	& \geq |r|.
\end{align}
Combining Eq.~\eqref{Eq_s_bound2} and Eq.~\eqref{Eq_r_bound} produces the desired bound $|s|^2 / 4  \geq |r|$.

\section{Distance bounds: part two}
\label{App_distances2}

Here we prove Eqs.~\eqref{breveDistances}. 

\subsection{First bound}

We begin with 
\begin{equation}
	 \breve{d}_0 \geq \mathrm{min}[ \tilde{d}_{-1} , \mathrm{max}[ \tilde{d}_{0} , \tilde{d}_{-1}^T ] , \tilde{d}_{0}^T  ] 
\end{equation}	  
and remark that the proof for $\breve{d}_{-1}^T $ will follow a similar fashion.  Recall that $\breve{d}_0$  is the  weight of the smallest vector $r$ such that $\breve{\delta}_{0}r = 0$ and $r \notin \im ( \breve{\delta}_{-1} )$. All $r$ can be decomposed as
\begin{equation}
r = \left( \begin{array}{c}
r_a \\
r_b \\
r_c \\
\end{array} \right) ,
\end{equation}
where  $\breve{\delta}_0 r =0$ requires that
\begin{align}
	\label{BreveBoundaryCond}
	(\id \otimes \tilde{\delta}_{-1}^T) r_b		 & =  (\tilde{\delta}_{-1} \otimes \id) r_a   ,  \\ \nonumber
	(\tilde{\delta}_{0} \otimes \id ) r_b     & =(\id \otimes \tilde{\delta}_{0}^T ) r_c   .
\end{align}
Taking the components of $r$ and reshaping into a matrices, the vector equations transform into matrix equations as follows
\begin{align}
	(\tilde{\delta}_{0} \otimes \id ) r_b      =(\id \otimes \tilde{\delta}_{0}^T ) r_c   & \implies \tilde{\delta}_{0} R_b  = R_c   \tilde{\delta}_{0}  ,\label{RcRbSwitch} \\
	(\id \otimes \tilde{\delta}_{-1}^T) r_b		  =  (\tilde{\delta}_{-1} \otimes \id) r_a 0   & \implies R_b \tilde{\delta}_{-1} = \tilde{\delta}_{-1} R_a \label{RaRbSwitch} .
\end{align}
Assuming $r$ is a non-trivial cycle, it follows that there must exist a nontrivial cocycle $w=w_a \oplus w_b \oplus w_c$ such that $w^T r=1$.  Furthermore, the cocycle can be assumed to be of the form $w=(e_a \otimes f_a) \oplus ( e_b \otimes f_b) \oplus ( e_c \otimes f_c )$ since the span of such vectors encompasses all nontrivial cocycles.

Therefore,  $w_a^Tr_a + w_b^T r_b + w_c^T r_c =1$ and at least one of these terms must equal 1 and there are four cases to consider
\begin{enumerate}
		\item $w_a^Tr_a=1$ and $w_b^T r_b = w_c^T r_c =0$, in which case we may assume $w= (e_a \otimes f_a) \oplus  0 \oplus 0 $;
		\item $w_c^Tr_c=1$ and $w_a^T r_a = w_b^T r_b =0$, in which case we may assume $w= 0  \oplus  0 \oplus (e_c \otimes f_c) $;
		\item $w_b^Tr_b=1$ and $w_a^T r_a = w_c^T r_c =0$, in which case we may assume $w=0 \oplus (e_b \otimes f_b) \oplus 0$;
		\item $w_a^Tr_a=w_b^T r_b = w_c^T r_c =1$; in which case we can find a new $w$ satisfying one of the above 3 cases.
\end{enumerate}
We again remind the reader that all vectors are column vectors.  Furthermore,  $\oplus$ is the direct product and when applied to columns vectors means that we stack the columns. 

We first consider case 1.  For $w= ( e_a \otimes f_a ) \oplus 0 \oplus 0 $ to be a cocycle requires that $\tilde{\delta}_{-1} f_a =0$.  Furthermore, the condition $w^T r = (e_a^T \otimes f_a^T) r_a =1$ in reshaped form becomes $e_a^T R_a f_a =1$.  We consider the vector $R_a f_a$, and find
\begin{align}
\tilde{\delta}_{-1} R_a f_a =  R_b \tilde{\delta}_{-1} f_a = 0, 
	\end{align}
where we have used Eq.~\eqref{RaRbSwitch} and $\tilde{\delta}_{-1} f_a=0$.  In other words, $R_a f_a \in \ker ( \tilde{\delta}_{-1}  )$.  However, $R_a f_a$ is non-zero otherwise it would be impossible to satisfy $e_a^T R_a f_a =1$.  It follows that $\tilde{d}_{-1} \leq |R_a f_a| $. Since $R_a f_a$ is formed from linear combinations of columns from $R_a$, we have $|R_a f_a| \leq |R_a|$ and hence $\tilde{d}_{-1} \leq |R_a|  $.  It follows that $\tilde{d}_{-1} \leq |r| $ in case 1. 

Next, we consider case 2. The proof method is essentially the same but we repeat for completeness. For $w= 0 \oplus 0 \oplus ( e_c \otimes f_c )$ to be a cocycle requires that $\tilde{\delta}_0^T e_c = 0$.  Furthermore, the condition  $w^T r = (e_c^T \otimes f_c^T) r_c =1$ in reshaped form becomes $e_c^T R_c f_c =1$.  We consider the vector $R_c^T e_c$, and find
\begin{align}
	\tilde{\delta}_{0}^T R_c^T e_c =  (R_c \tilde{\delta}_{0})^T e_c =  ( \tilde{\delta}_{0} R_b)^T e_c  = R_b^T \tilde{\delta}_{0}^T e_c = 0, 
\end{align}
where we have used Eq.~\eqref{RcRbSwitch} and $\tilde{\delta}_0^T e_c = 0$.   In other words, $R_c^T e_c \in \ker ( \tilde{\delta}_{0}^T  )$.  However, $R_c^T e_c$ is non-zero otherwise it would be impossible to satisfy $e_c^T R_c f_c =1$.  It follows that $\tilde{d}_{0}^T \leq |R_c^T e_c|$. Since $R_c^T e_c$ is formed from linear combinations of rows from $R_c$, we have $|R_c^T e_c| \leq |R_c|$ and hence $\tilde{d}_{0}^T \leq  |R_c|$.  It follows that $\tilde{d}_{0}^T \leq |r| $ in case 2. 

Next, we consider case 3 then $w=0 \oplus (e_b \otimes f_b) \oplus 0 $.  Furthermore, the condition  $w^T r = (e_b^T \otimes f_b^T) r_b =1$ in reshaped form becomes $e_b^T R_b f_b =1$.  The proof is slightly different from the above two cases.  The cocycle conditions now tells us that both $\tilde{\delta}_{-1}^T e_b = 0$ and $\tilde{\delta}_{0} f_b = 0$.  We have
\begin{align}
\tilde{\delta}_{0}  R_b f_b & =  R_c   \tilde{\delta}_{0} f_b  = 0  , \\
\tilde{\delta}_{-1}^T	R_b^T e_b & = (	R_b \tilde{\delta}_{-1})^T  e_b =  (\tilde{\delta}_{-1} R_a)^T  e_b  = R_a^T \tilde{\delta}_{-1}^T e_b = 0 ,
\end{align}
 where we have used $ \tilde{\delta}_{0} f_b=0 $ and  $\tilde{\delta}_{-1}^T e_b = 0$  as asserted earlier. Furthermore, $R_b f_b \notin \im (\tilde{\delta}_{-1})  $ since otherwise $R_b f_b = \tilde{\delta}_{-1} u$ for some $u$ and then $e_b^T R_b f_b =  e_b^T  \tilde{\delta}_{-1} u = (\tilde{\delta}^T_{-1} e_b  )^T u$.  However, since $\tilde{\delta}^T_{-1} e_b=0$ this would entail $e_b^T R_b f_b = 0$ which is a contradiction and so we must have $R_b f_b \notin \im (\tilde{\delta}_{-1})  $.  Similarly, one has that  $R_b^T e_b  \notin \im (\tilde{\delta}_{0}^T)  $ otherwise $R_b^T e_b = \tilde{\delta}_{-1} v$ for some $v$ which would again lead to the  contradiction $e_b^T R_b f_b = 0$  when combined with the fact that $\tilde{\delta}_{0} f_b = 0$.  Combining $R_b f_b \in \ker (\tilde{\delta}_{0} )$ and  $R_b f_b \notin \im (\tilde{\delta}_{-1})$ entails that $R_b f_b$ is a nontrivial cycle and so $\tilde{d}_{0} \leq |R_b f_b| $.  Since $R_b f_b$ is formed from linear combinations of columns from $R_b$, we have $|R_b f_b| \leq |R_b|$ and hence $\tilde{d}_{0} \leq  |R_b|$. Similarly, combining $R_b^T e_b \in \ker( \tilde{\delta}_{-1}^T )$ and $R_b^T e_b  \notin \im (\tilde{\delta}_{0}^T)$  leads to  $\tilde{d}_{-1}^T \leq |R_b| $.  This suffices to prove that in case 3 we have $|r| \geq \mathrm{max}[ \tilde{d}_{0} , \tilde{d}_{-1}^T ]$. 
 
Since any one of the three cases may hold, we must take the minimum over the three cases.  This yields the distance lower bound on $\breve{d}_0$.

\subsection{Second bound}

Here we prove 
\begin{equation}
	\breve{d}_{1}  \geq \mathrm{min}[ \tilde{d}_{0} , \tilde{d}_{-1}^T   ] ,
\end{equation}	  
and remark that the proof for $\breve{d}_{-2}^T $ will follow a similar fashion.  Let $s =  s_a \oplus s_b  \in \breve{C}_1$ be a minimal distance nontrivial cycle for $\breve{\delta}_{1}$.  From $\breve{\delta}_{1} s = 0$ we may infer
\begin{equation}
		(\tilde{\delta}_0 \otimes \id) s_a = (\id \otimes \tilde{\delta}_{-1}^T) s_b  .
\end{equation}	
Since $s$ is a nontrivial cycle, there must exist a nontrivial cocycle $w= w_a \oplus w_b $ such that $w^T s =1$.  There are two possible cases
\begin{enumerate}
	\item $w_a^T s_a=1$ and $w_b^T s_b =0$, in which case we may assume $w= ( e_a \otimes f_a  ) \oplus 0 $;
	\item $w_b^T s_b=1$ and $w_a^T s_a=0$, in which case we may assume $w=0 \oplus ( e_b \otimes f_b )$;
\end{enumerate}	
For case 1, since $w$ is a cocycle $\breve{\delta}_{1}^T w = 0$ and so both $\tilde{\delta}_{-1}^T e_a =0$ and  $\tilde{\delta}_{-1} f_a =0$.  However, $e_a \notin \im (\tilde{\delta}_0)$ otherwise $w$ would be a trivial cocycle.  As in other proofs, we now reshape into matrix equations
\begin{align}
 w^T s =1 & \implies	e_a^T S_a f_a  = 1 \label{SWIdentity}  \\
(\tilde{\delta}_0 \otimes \id) s_a = (\id \otimes \tilde{\delta}_{-1}^T) s_b  & \implies	\tilde{\delta}_0 	S_a  = S_b \tilde{\delta}_{-1} \label{SabIdentity}. 
\end{align}	
Therefore, 
\begin{align}
\tilde{\delta}_{0} (S_a f_a) & = S_b \tilde{\delta}_{-1} f_a = 0 
\end{align}	
where we have used Eq.~\eqref{SabIdentity} and $ \tilde{\delta}_{-1} f_a =0$.  In other words, $S_a f_a \in \ker ( \tilde{\delta}_{0}  )$.  However,  $S_a f_a \notin \im ( \tilde{\delta}_{-1}  )$ otherwise there would exist a $u$ such that $S_a f_a = \tilde{\delta}_{-1}  u$ and then $e_a^T S_a f_a = e_a^T \tilde{\delta}_{-1}  u = 0$ by virtue of $\tilde{\delta}_{-1}^T e_a =0$.  This is in contradiction with Eq.~\eqref{SWIdentity} and so $S_a f_a$ is a nontrivial cycle of $\tilde{\delta}_{0}$ and must satisfy $\tilde{d}_{0} \leq |S_a f_s| $.  It follows that $\tilde{d}_{0} \leq |S_a| \leq |s|$.  

For case 2, a similar proof entails that $\tilde{d}_{-1}^T  \leq |S_b| \leq |s|$.  Since either case may hold the distance is given by the minimum of these two quantities.

\section{Partial soundness}
\label{App_Partial_coexpand}

\begin{figure}
	\includegraphics{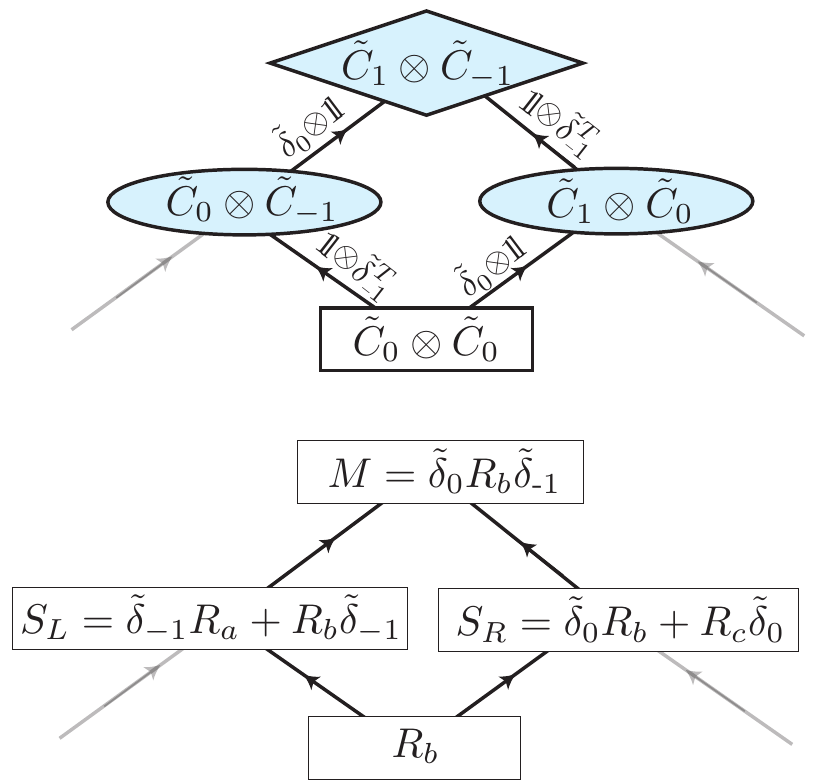}
	\caption{Top: the relevant subgraph of Fig.~\ref{Fig_DoubleHP} reproduced here for convenient reference. Bottom:  the relations between different reshaped matrices as given in Eqs.~\eqref{ReshapeRelations1},~\eqref{ReshapeRelations2} and~\eqref{ReshapeRelations3}.  Here we draw the readers attention to how these two figures are connected.  For instance $r_b$ is an element of vector space $\tilde{C}_0 \otimes \tilde{C}_0$, which is then reshaped into  $R_b$. }
	\label{Fig_DoubleHPSubFig}
\end{figure}

\begin{figure*}[h] \framebox{
		\begin{minipage}{.6\textwidth}
			\raggedright
			{\bf Input:} A set of matrices $R_b$, $S_L$, $S_R$, $\tilde{\delta}_0$ and $\tilde{\delta}_{-1}$, with relationships defined in main text. \\
			{\bf Output:} A new transformed $R_b'$ such that $\tilde{\delta}_0 R'_b \tilde{\delta}_{-1}=\tilde{\delta}_0 R_b \tilde{\delta}_{-1}$ and furthermore $R'_b$ satisfies a set of constraints on its column and row support.
			\begin{enumerate}
				\item While $\mathrm{rowsupp}(R_b \tilde{\delta}_{-1}) -  \mathrm{rowsupp}(S_L)$ is nonempty
				\begin{enumerate}
					\item $i \leftarrow SAMPLE [ \mathrm{rowsupp}(R_b \tilde{\delta}_{-1}) -  \mathrm{rowsupp}(S_L) ] $;
					\item $v^T \leftarrow$ the $i^{\mathrm{th}}$ row of $R_b$;
					\item $r^T \leftarrow $ the $i^{\mathrm{th}}$ row of $R_b \tilde{\delta}_{-1}$;
					\item $j \leftarrow SAMPLE [ \mathrm{colsupp}(r^T) ]$;
					\item $c \leftarrow $ the $j^{\mathrm{th}}$ column of $R_b \tilde{\delta}_{-1}$;
					\item $c' \leftarrow $ the $j^{\mathrm{th}}$ column of $S_L$;
					\item $w \leftarrow c + c'$;
					\item $R_b \leftarrow R_b + w v^T$;
				\end{enumerate}
				\item While $\mathrm{colsupp}(R_b \tilde{\delta}_{-1}) -  \mathrm{colsupp}(S_L)$ is nonempty
				\begin{enumerate}
					\item $j \leftarrow SAMPLE [ \mathrm{colsupp}(R_b \tilde{\delta}_{-1}) - \mathrm{colsupp}(S_L) ] $;
					\item $c \leftarrow $ the $j^{\mathrm{th}}$ column of $R_b \tilde{\delta}_{-1}$;
					\item $k \leftarrow SAMPLE [ \mathrm{rowsupp}(R_b) \cap \mathrm{rowsupp}(c)  ]$
					\item $v^T \leftarrow$ the $k^{\mathrm{th}}$ row of $R_b$;
					\item $R_b \leftarrow R_b + c v^T$;
				\end{enumerate}
				\item While $\mathrm{rowsupp}(R_b) - \mathrm{rowsupp}(R_b \tilde{\delta}_{-1}) $ is nonempty
				\begin{enumerate}
					\item $j \leftarrow SAMPLE [ \mathrm{rowsupp}(R_b) - \mathrm{rowsupp}(R_b \tilde{\delta}_{-1}) ] $;
					\item $j^{\mathrm{th}}$ row of $R_b$ $\leftarrow (0,0,\ldots,0)$ 
				\end{enumerate}
				\item While $\mathrm{colsupp}( \tilde{\delta}_{0} R_b )-  \mathrm{colsupp}(S_R)$ is nonempty
				\begin{enumerate}
					\item $i \leftarrow SAMPLE [ \mathrm{colsupp}(\tilde{\delta}_{0} R_b) -  \mathrm{colsupp}(S_R) ] $;
					\item $v \leftarrow$ the $i^{\mathrm{th}}$ column of $R_b$;
					\item $r \leftarrow $ the $i^{\mathrm{th}}$ column of $\tilde{\delta}_{0} R_b$;
					\item $j \leftarrow SAMPLE [ \mathrm{rowsupp}(r) ]$;
					\item $c \leftarrow $ the $j^{\mathrm{th}}$ row of $\tilde{\delta}_0 R_b$;
					\item $c' \leftarrow $ the $j^{\mathrm{th}}$ row of $S_R$;
					\item $w^T \leftarrow c + c'$;
					\item $R_b \leftarrow R_b + v w^T$;
				\end{enumerate}
				\item While $\mathrm{rowsupp}(\tilde{\delta}_{0} R_b ) -  \mathrm{rowsupp}(S_R)$ is nonempty
				\begin{enumerate}
					\item $j \leftarrow SAMPLE [ \mathrm{rowsupp}(\tilde{\delta}_{0} R_b ) -  \mathrm{rowsupp}(S_R) ]$;
					\item $v^T \leftarrow $ the $j^{\mathrm{th}}$ row of $\tilde{\delta}_{0} R_b $;
					\item $k \leftarrow SAMPLE [ \mathrm{colsupp}(R_b) \cap \mathrm{colsupp}(v^T)  ]$
					\item $c \leftarrow$ the $k^{\mathrm{th}}$ column of $R_b$;
					\item $R_b \leftarrow R_b + c v^T$;
				\end{enumerate}
				\item While $\mathrm{colsupp}(R_b) - \mathrm{colsupp}(\tilde{\delta}_{0} R_b) $ is nonempty
				\begin{enumerate}
					\item $j \leftarrow SAMPLE [ \mathrm{colsupp}(R_b) - \mathrm{colsupp}( \tilde{\delta}_{0} R_b ) ] $;
					\item $j^{\mathrm{th}}$ column of $R_b$ $\leftarrow (0,0,\ldots,0)^T$ 
				\end{enumerate}
			\end{enumerate} 
			{\bf Return:} $R_b$.
	\end{minipage}  }
	\caption{A partial decoder.  Certain choices are arbitrary and so we use $SAMPLE[\ldots]$ to mean randomly sample (or use any other criteria) to select one element from a set.  For an example of step 1 see transform 1 of toy example 3 in Fig.~\ref{Fig_SecondSoundnessToy3}. For an example of step 2 see transform 1 of toy example 1 in Fig.~\ref{Fig_SecondSoundness}.  For an example of step 3 see transform 2 of toy example 1 in Fig.~\ref{Fig_SecondSoundness}. For an example of step 5 see transform 1 of toy example 2 in Fig~\ref{Fig_SecondSoundnessToy2}.  See the supplementary material for a Mathematica implementation of this partial decoder.}
	\label{MyPartialDecoder}
\end{figure*}

Here we prove Lem.~\ref{Lem_MiniCoexpand}, which is a major technical component of Thm.~\ref{Lem_MiniCoexpand}.  We are working towards a low-weight solution of 
\begin{align} 
\label{Eq_Mcondition}
m = (\tilde{\delta}_0  \otimes \tilde{\delta}_{-1}^T)  r_b .
\end{align}	
So far we only know that there must be at least one $r_b$ satisfying this equation.  We proceed by looking for other $r_b$ consistent with Eq.~\eqref{Eq_Mcondition} that have a low weight and other additional properties.  At this point it is convenient to reshape our vectors into matrices (recall App.~\ref{App_Matrix_reshaping}) and the previous equations become
\begin{align}
	\label{ReshapeRelations1}
		S_L & = 		\tilde{\delta}_{-1}  R_a + R_b  \tilde{\delta}_{-1} , \\ 	\label{ReshapeRelations2}
        S_R & = 		\tilde{\delta}_0 R_b +  R_c \tilde{\delta}_{0} , \\ 	\label{ReshapeRelations3}
		M & =   \tilde{\delta}_0 S_L =  S_R  \tilde{\delta}_{-1} = \tilde{\delta}_0  R_b  \tilde{\delta}_{-1}. 
\end{align}
As a visual aid to understanding these equations we provide Fig.~\ref{Fig_DoubleHPSubFig}.

Notice that if $R_b$ has any columns in the kernel of $\tilde{\delta}_0$, these can be removed without changing $M$.  Similarly, if $R_b$ has any rows in the kernel of $\tilde{\delta}_{-1}^T$, these can be removed without changing $M$.  So we see there are transforms that preserve $M$ but remove elements from $R_b$.  

\begin{figure*}[t]
\includegraphics[width=0.9\textwidth]{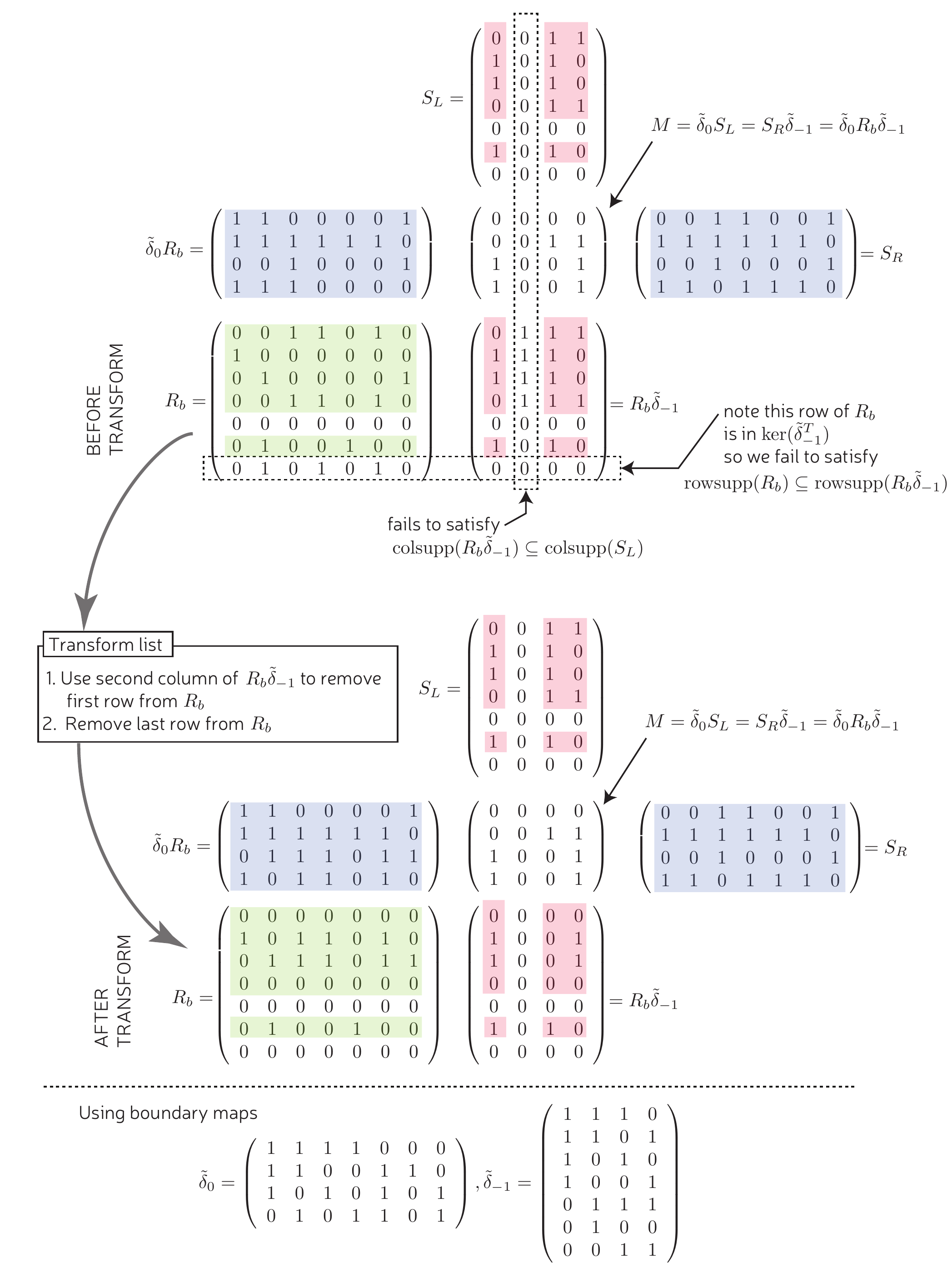}
\caption{Toy example 1 showing the form of an initial $R_b$ matrix before any transformations have been performed. The matrix $\tilde{\delta}_0$ was not generated by the homological product but otherwise all features are correct.  An actual homological product example would be too large to be instructive and furthermore the partial soundness proof does not use any such properties.  The goal is to transform $R_b$ such that $M$ is unchanged, but after the transform $R_b$, $R_b \tilde{\delta}_{-1}$ and  $\tilde{\delta}_{0} R_b $ are only supported within the highlighted boxes.  The highlighted boxes are themselves derived from the column and row support of $S_L$ and $S_R$ that are fixed.}
\label{Fig_SecondSoundness}
\end{figure*}

\begin{figure*}[t]
\includegraphics[width=0.9\textwidth]{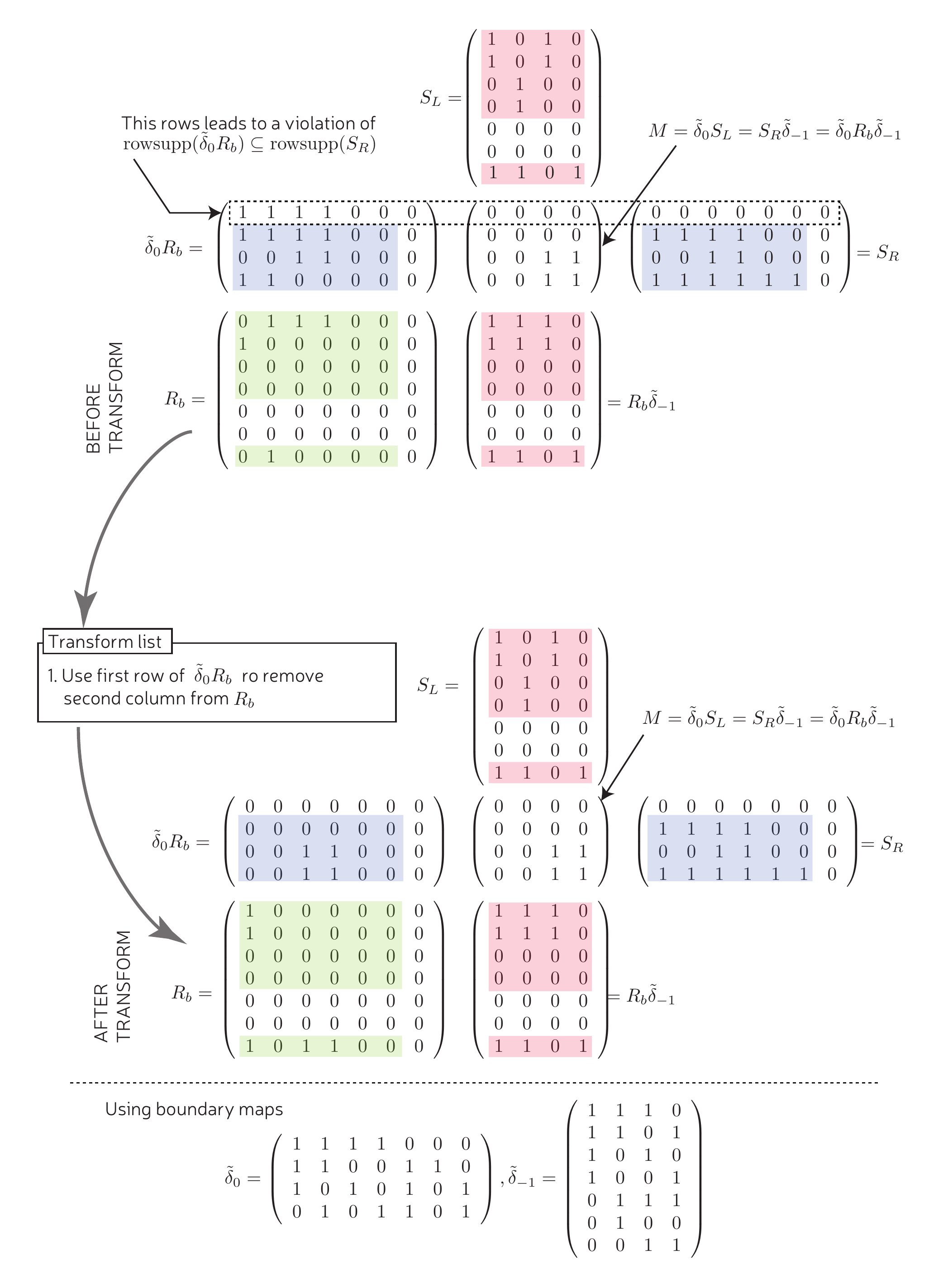}
\caption{Toy example 2 showing the form of an initial $R_b$ matrix before any transformations have been performed.  All $\delta$ boundary maps are the same as in toy example 1 shown in Fig.~\ref{Fig_SecondSoundness}.}
\label{Fig_SecondSoundnessToy2}
\end{figure*}

\begin{figure*}[t]
\includegraphics[width=0.9\textwidth]{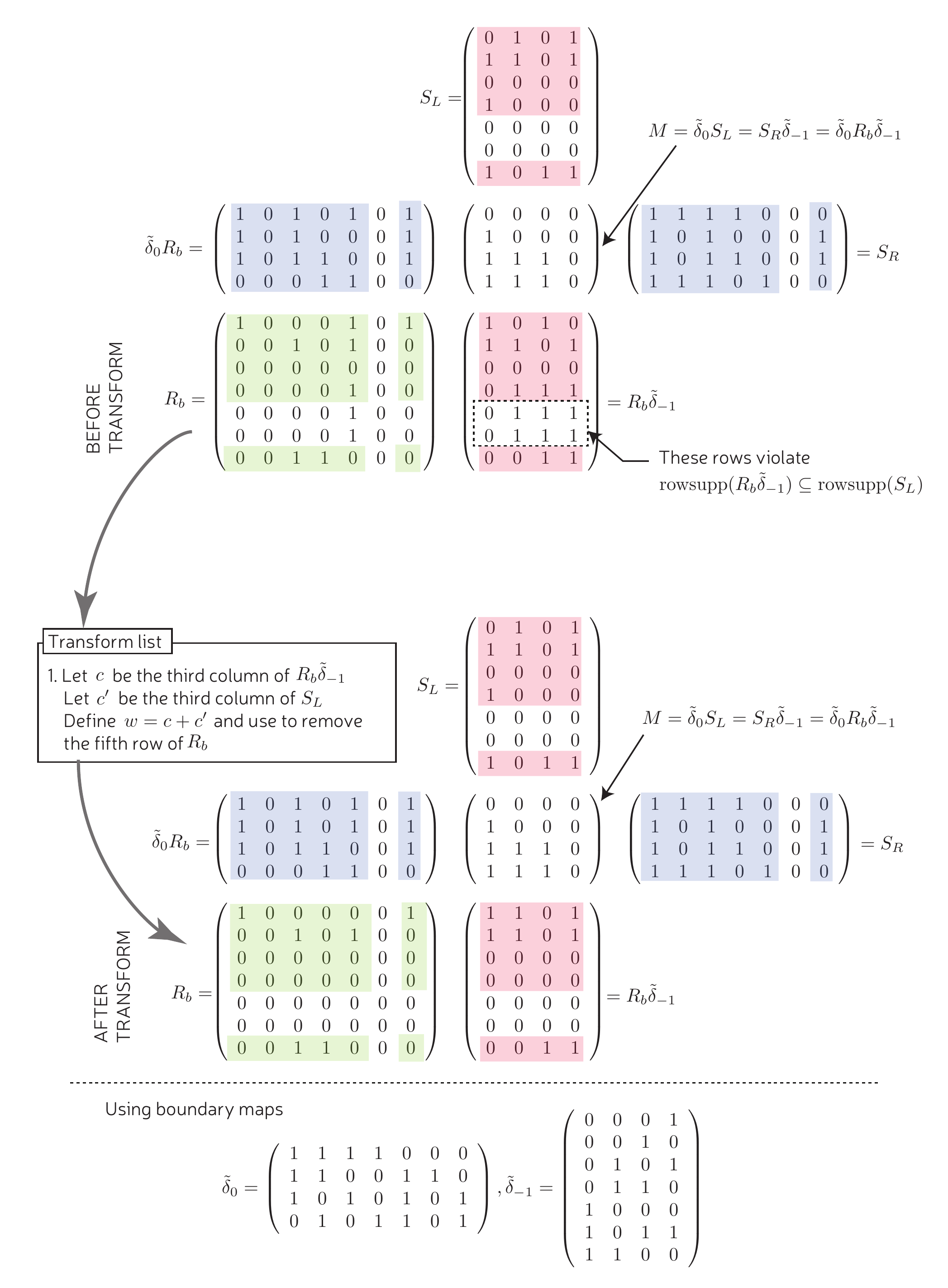}
\caption{Toy example 3.  Note that in this example we use a different boundary map $\tilde{\delta}_{-1}$ just for the sake of variety.}
\label{Fig_SecondSoundnessToy3}
\end{figure*}

Our proof is essentially a decoder for $R_b$.  This is a partial decoder as it requires an initial guess for $R_b$ and does not solve for $R_a$ and $R_c$.  We describe the decoder in pseudocode in Fig.~\ref{MyPartialDecoder} and give toy examples of it's implementation in Figs.~\ref{Fig_SecondSoundness}, ~\ref{Fig_SecondSoundnessToy2} and~\ref{Fig_SecondSoundnessToy3}.  The rest of this section will discuss the possible transforms of $R_b$ and then an analysis  of the partial decoder.

\textbf{\textit{Support inclusions.-}} Simple matrix algebra (recall notation from App.~\ref{App_Matrix_support}) leads  to the inclusions 
\begin{align}
\mathrm{rowsupp} ( R_b  \tilde{\delta}_{-1} )   & \subseteq \mathrm{rowsupp} ( R_b   ) , 	\\
\mathrm{colsupp} ( 	\tilde{\delta}_0 R_b  ) & \subseteq \mathrm{colsupp} ( 	 R_b  ) , \\
\mathrm{colsupp} ( M  ) & \subseteq \mathrm{colsupp} ( 	 S_L  ) , \\
\mathrm{rowsupp} ( M  ) & \subseteq \mathrm{rowsupp} ( 	 S_R  ) .
\end{align}
If we inspect our toy examples (Figs.~\ref{Fig_SecondSoundness}, ~\ref{Fig_SecondSoundnessToy2} and~\ref{Fig_SecondSoundnessToy3}) we see that these are indeed satisfied before any transformations are performed.

The goal of the partial decoder is to perform a series of transforms such that $M$ is preserved and the final output $R_b$ satisfies the following:
\begin{align} 	\label{NewReshapeRelations1}
		\mathrm{rowsupp} ( R_b  \tilde{\delta}_{-1} )   & \subseteq \mathrm{rowsupp} ( S_L   ) 	 , \\  	\label{NewReshapeRelations2}
		\mathrm{colsupp} ( R_b  \tilde{\delta}_{-1} )   & \subseteq \mathrm{colsupp} ( S_L   ) 	,\\ 	\label{NewReshapeRelations3}
		\mathrm{rowsupp} ( R_b  \tilde{\delta}_{-1} )   & = \mathrm{rowsupp} ( R_b   ) 	,\\			 	\label{NewReshapeRelations4}
	\mathrm{colsupp} ( 	\tilde{\delta}_0 R_b  ) & \subseteq \mathrm{colsupp} ( 	S_R  ) , \\ 	\label{NewReshapeRelations5}
	\mathrm{rowsupp} ( 	\tilde{\delta}_0 R_b  ) & \subseteq \mathrm{rowsupp} ( 	S_R  )  , \\ 	\label{NewReshapeRelations6}
	\mathrm{colsupp} ( 	\tilde{\delta}_0 R_b  ) & = \mathrm{colsupp} ( 	 R_b  )  , 
\end{align}	
The partial decoder goes through 6 while loops with each loop aiming to enforce one of these conditions.  In each case, the idea is that if the condition is violated this enables us to perform some $M$ preserving transformation that removes columns or rows from $R_b$.

\textbf{\textit{Overview of $R_b$ transforms.-}}  Next, we give a very general account of how we may transform $R_b$ while preserving $M$.  Given a column vector $c$ such that $\tilde{\delta}_0 c = 0$, we may add $c$ to any of the columns in $R_b$ and $M$ will not change.  Furthermore, if $\mathrm{rowsupp} ( R_b  )$ and   $\mathrm{rowsupp} ( c )$ have any elements in common, then we can perform a transformation that removes one row from $R_b$.  Also note that if $c$ is itself a column vector of $R_b $ then it is trivially the case that they share row support in common. This is similar to the intersecting argument encountered in the proof in App.~\ref{App_Coexpand1}.  Let us again illustrate by example.  Suppose
\begin{align}
	 R_b = \left( \begin{array}{cccc}
	 	1 & 1 & 1 & 0 \\
	 	1 & 0 & 1 & 0 \\
	    0 & 0 & 0 & 0 \\
	    0 & 1 & 1 & 1
	 	\end{array} \right) \mbox{ and } & c = \left( \begin{array}{c}
	 	1 \\
	 	1 \\
	 	1 \\
	 	0  \\
 	\end{array} \right) ,
\end{align}	 
so that $\mathrm{rowsupp} ( R_b  ) = \{ 1,2,4 \}$ and $\mathrm{rowsupp} ( c  ) = \{ 1,2,3 \}$.  We see that both supports share $1$ and $2$ in common and so either row could be removed.  For example, to remove row 1 we add $c$ to columns 1, 2 and 3, yielding
\begin{align}
	R_b = \left( \begin{array}{cccc}
		0 & 0 & 0 & 0 \\
		0 & 1 & 0 & 0 \\
		1 & 1 & 1 & 0 \\
		0 & 1 & 1 & 1
	\end{array} \right)  ,
\end{align}	
where row 1 is now trivial.   Note that while row 1 has been removed, $\mathrm{rowsupp} ( R_b )$ now includes row 3, so the total number of supported rows has not decreased.   Note the $\mathrm{colsupp} ( R_b  )$ has not gained any new elements.

Let us now consider another example
\begin{align}
	R_b = \left( \begin{array}{cccc}
		1 & 1 & 1 & 0 \\
		1 & 0 & 1 & 0 \\
		0 & 1 & 1 & 1 \\
		0 & 0 & 0 & 0
	\end{array} \right) \mbox{ and } & c = \left( \begin{array}{c}
		1 \\
		1 \\
		1 \\
		0  \\
	\end{array} \right) ,
\end{align}	 
which is similar to the earlier example except now $\mathrm{rowsupp} ( R_b  )$ is equal to $\mathrm{rowsupp} ( c  )$.  Consequently, when we  use $c$ to remove row 1 we obtain
\begin{align}
	R_b = \left( \begin{array}{cccc}
		0 & 0 & 0 & 0 \\
		0 & 1 & 0 & 0 \\
		1 & 0 & 0 & 1 \\
		0 & 0 & 0 & 0
	\end{array} \right)  ,
\end{align}
so row 1 has been removed but also the total number of rows has decreased.   Note again that $\mathrm{colsupp} ( R_b  )$ has not gained any new elements.

More generally, we have that
\begin{claim}[Row removal]
	\label{RowRemove}
Let $c$ be a column vector such that $\tilde{\delta}_{0}c=0$ and let $v^T$ be the $j^{\mathrm{th}}$ row vector of $R_b$ where $j \in \mathrm{rowsupp} ( R_b  ) \cap \mathrm{rowsupp} ( c )$. Then the transform $R_b \rightarrow R_b'=R_b + c v^T$  satisfies the following
\begin{enumerate}
	 \item the transform will preserve $M$;
	 \item  the new $R'_b$ will have row support in $\mathrm{rowsupp} ( 	 R_b  )  \cup \mathrm{rowsupp} ( c )  - \{ j \} $.  If one further has that $\mathrm{rowsupp} ( c ) $ is contained within $\mathrm{rowsupp} ( 	 R_b  ) $ then the number of rows has strictly decreased.
	 \item  the new $R'_b$ will have column support within the original $\mathrm{colsupp} ( 	 R_b  )$. 
\end{enumerate}
\end{claim}
Similarly, 
\begin{claim}[Column removal]
	\label{ColRemove}
Let $v^T$ be a row vector such that $v^T \tilde{\delta}_{-1}=0$ and let $c$ be the $j^{\mathrm{th}}$ column vector of $R_b$ where $j \in \mathrm{colsupp} ( R_b  ) \cap \mathrm{colsupp} ( v )$. Then the transform $R_b \rightarrow R_b'=R_b + c v^T$  satisfies the following
\begin{enumerate}
	\item the transform will preserve $M$;
	\item  the new $R'_b$ will have column support in $\mathrm{colsupp} ( 	 R_b  )  \cup \mathrm{colsupp} ( v^T )  - \{ j \} $.  If one further has that $\mathrm{colsupp} ( v^T ) $ is contained within $\mathrm{colsupp} ( 	 R_b  ) $ then the number of columns has strictly decreased.
	\item  the new $R'_b$ will have row support within the original $\mathrm{rowsupp} ( 	 R_b  )$. 
\end{enumerate}
\end{claim}
We now proceed to use these ideas in the following way.  

\textbf{\textit{While loop 1}.-} This iteratively reduces the number of elements in  $\mathrm{rowsupp} ( R_b  \tilde{\delta}_{-1} )    \cup \mathrm{rowsupp} ( S_L   ) $ until we have $\mathrm{rowsupp} ( R_b  \tilde{\delta}_{-1} )    \subseteq \mathrm{rowsupp} ( S_L   ) $.   Whenever this inclusion is false, there exists at least one column, say $c$, of  $R_b  \tilde{\delta}_{-1}$ such that  $ \mathrm{rowsupp} ( c  )$ is not a subset of $\mathrm{rowsupp} ( S_L   ) $.  Furthermore, if $c$ is the $j^{\mathrm{th}}$ column of $R_b  \tilde{\delta}_{-1}$ let $c'$ be the $j^{\mathrm{th}}$ column of $S_L$.   We must have that $c' \neq c$ otherwise $ \mathrm{rowsupp} ( c  )$ would be a subset of $\mathrm{rowsupp} ( S_L   ) $. Since $ \tilde{\delta}_{0} R_b  \tilde{\delta}_{-1}=  \tilde{\delta}_{0} S_L$ we must have that these matrices are equal on the $j^{\mathrm{th}}$ column and so $ \tilde{\delta}_{0} c =  \tilde{\delta}_{0} c'$.  Therefore, the vector $w=c' - c$ satisfies the following properties:
\begin{enumerate}
	\item  $w \neq 0$ which follows from $c \neq c'$;
	\item $w \in \ker ( \tilde{\delta}_{0} )$ which follows from $ \tilde{\delta}_{0} c =  \tilde{\delta}_{0} c'$;
	\item $\mathrm{rowsupp} ( w ) \subseteq \mathrm{rowsupp} (R_b  \tilde{\delta}_{-1} ) \cup  \mathrm{rowsupp} (  S_L )$ which follows from $\mathrm{rowsupp} ( w ) \subseteq \mathrm{rowsupp} ( c ) \cup  \mathrm{rowsupp} (  c' )$.
	\item $\mathrm{rowsupp} ( w ) \cap \mathrm{rowsupp} (R_b )$ is non-empty, because $c$ (and hence $w$) has row support outside  $ \mathrm{rowsupp} (  S_L )$.
\end{enumerate}
Therefore, we can (by virtue of claim~\ref{RowRemove}) use column vector $w$ to remove a row from $R_b$. The row removal process is possible for any row in $\rowsupp(R_b) \cap \rowsupp(w)$.  However, we want the final row support to be within $\mathrm{rowsupp} ( S_L   )$ and so from the set of possible rows we choose one outside the set $\mathrm{rowsupp} ( S_L   )$.  In practice, the partial decoder pseudocode make this row selection the first task.   Therefore, the set  $(\mathrm{rowsupp} (R_b  \tilde{\delta}_{-1} ) \cup  \mathrm{rowsupp} (  S_L ))-\mathrm{rowsupp} (  S_L )$  strictly decreases in size.  Note also that $\mathrm{colsupp} (R_b)$ will not increase (by clause 2 of claim~\ref{RowRemove}). For an example, see transform 1 of toy example 3 in Fig.~\ref{Fig_SecondSoundnessToy3}.

\textbf{\textit{While loop 2}.-}  This iteratively reduces the number of rows in $R_b$ until we have $\mathrm{colsupp} ( R_b  \tilde{\delta}_{-1} )   \subseteq \mathrm{colsupp} ( S_L   ) $.   First note that if $R_b  \tilde{\delta}_{-1}$ has any nonzero columns outside $\mathrm{colsupp} ( S_L   )$, the column must be in the kernel of $\tilde{\delta}_0$.  To prove this, note that if the offending column was outside $\ker(\tilde{\delta}_0)$ then  $\mathrm{colsupp} ( \tilde{\delta}_{0} R_b  \tilde{\delta}_{-1} )$ would be strictly larger than $\mathrm{colsupp} ( \tilde{\delta}_{0} S_L)$ which contradicts $\tilde{\delta}_{0} R_b  \tilde{\delta}_{-1} = \tilde{\delta}_{0} S_L$.  Since the column is in $\ker(\tilde{\delta}_0)$ and within $\mathrm{colsupp} ( R_b  \tilde{\delta}_{-1} )$, its presence allows us (by virtue of claim~\ref{RowRemove})  to remove a row from $R_b$.  It is crucial that after each iteration of the loop,  the column support of $R_b$ strictly decreases (by clause 2 of claim~\ref{RowRemove}), which entails that the while loop must terminate after a finite number of iterations.  It is important to comment on what we do not show here; we do not show that each iteration strictly removes columns from $\mathrm{colsupp} ( R_b  \tilde{\delta}_{-1} )$ until it is contained in $\mathrm{colsupp} ( S_L   )$.  Rather the number of rows in $R_b$ are strictly decreased and this process cannot continue without end, so the while loop termination criteria must be satisfied within a finite number of rounds. To be precise, the while loop must terminate, since either (1) after a finite number of loops we obtain some nonzero $R_b$ such that $\mathrm{colsupp} ( R_b  \tilde{\delta}_{-1} )   \subseteq \mathrm{colsupp} ( S_L   ) $; or (2) after a finite number of iterations all rows will be removed from $R_b$, so that $R_b=0$, and then $\mathrm{colsupp} ( R_b  \tilde{\delta}_{-1} ) = \mathrm{colsupp} ( 0 ) = \emptyset  $  is trivially true.   Again $\mathrm{colsupp} (R_b)$ will not increase. For an example, see transform 1 of toy example 1 in Fig.~\ref{Fig_SecondSoundness}. 

\textbf{\textit{While loop 3}.-} This iteratively reduces the number of rows in $R_b$ until $	\mathrm{rowsupp} ( R_b  \tilde{\delta}_{-1} )  = \mathrm{rowsupp} ( R_b   ) $.  This is a fairly straightforward step, since the offending rows must be in the kernel of  $\tilde{\delta}_{-1}^T$  they can just be simply removed.  Removing rows from $R_b$ leads to rows being removed from $R_b \tilde{\delta}_{-1}$ and the condition established in the previous while loop (that $\mathrm{colsupp} ( R_b  \tilde{\delta}_{-1} )$) will remain true.  For an example, see transform 2 of toy example 1 in Fig.~\ref{Fig_SecondSoundness}. 

\textbf{\textit{While loop 4}.-} This is similar to while loop 1, except with roles of rows and columns switched and applied to different matrices.  Here we reduce the number of elements in  $\mathrm{colsupp} (  \tilde{\delta}_{0} R_b  )    \cup \mathrm{colsupp} ( S_R   ) $ until we have $\mathrm{colsupp} ( 	\tilde{\delta}_0 R_b  )  \subseteq \mathrm{colsupp} ( S_R  )$, making use of  claim~\ref{ColRemove}.  Since the process does not introduce any new elements into $\mathrm{rowsupp} ( R_b   ) $, the previously established conditions will continue to hold true.

\textbf{\textit{While loop 5}.-} This is similar to while loop 2, except with roles of rows and columns switched and applied to different matrices and making use of claim~\ref{ColRemove}. For an example of step 5 see transform 1 of toy example 2 in Fig.~\ref{Fig_SecondSoundnessToy2}.

\textbf{\textit{While loop 6}.-} This is similar to while loop 3, except with roles of rows and columns switched and applied to different matrices. 

\textbf{\textit{Analysis}.-} The above process will terminate because the column and row support of $R_b$ is being gradually reduced. By repeating the above transformations until the process terminates, we ensure that $R_b  \tilde{\delta}_{-1}$ has row and column support strictly within that of $S_L$.    Therefore, the combination $S_L - R_b  \tilde{\delta}_{-1}$ also has row and column support strictly within that of $S_L$.   We can infer that $S_L - R_b  \tilde{\delta}_{-1} = \sum_i \alpha_i \otimes \hat{a}_i$ where $\alpha_i$ are the column vectors.  Since $S_L$ has at most $|S_L|$ columns, there can be at most $|S_L|$ nonzero $\alpha_i$.  Since $S_L$ has at most $|S_L|$ rows, each $\alpha_i$ has weight at most $|S_L|$.  This proves the small $|S_L|$ remainder property of our lemma (see property 3).  The small $|S_R|$ remainder property holds by a similar fashion (see property 4).  Furthermore, combining Eq.~\eqref{NewReshapeRelations1} and Eq.~\eqref{NewReshapeRelations3}, we conclude that the final $R_b$ has fewer rows than $S_L$ and so no more than $|S_L|$ rows.  Similarly, we deduce that the final $R_b$ has fewer columns than $S_R$ and so no more than $|S_R|$ rows.  Since the nonzero values of $R_b$ are contained within a submatrix of size $|S_L|$ by $|S_R|$, we know $|R_b| \leq |S_L| \cdot |S_R|$.  This proves property 2 of the lemma.  It should be clear that property 1 holds because the value of $M$ was initially correct and has been preserved through all transformations.
 

\end{document}